\documentclass[fleqn,10pt]{wlscirep}
\usepackage[utf8]{inputenc}
\usepackage[T1]{fontenc}
\usepackage{array,booktabs}
\usepackage{siunitx}
\usepackage{multirow,tabularx}
\title{Deep Learning Predicts Total Knee Replacement from Magnetic Resonance Images}
\author[1,2]{Aniket A. Tolpadi}
\author[2]{Jinhee J. Lee}
\author[2]{Valentina Pedoia}
\author[2,*]{Sharmila Majumdar}

\affil[1]{University of California, Berkeley, Department of Bioengineering, Berkeley, 94720, USA}
\affil[2]{University of California, San Francisco, Department of Radiology and Biomedical Imaging, San Francisco, 94158, USA}

\affil[*]{Sharmila.Majumdar@ucsf.edu}

\newcolumntype{P}[1]{>{\centering\arraybackslash}p{#1}}

\begin{abstract}
Knee Osteoarthritis (OA) is a common musculoskeletal disorder in the United States. When diagnosed at early stages, lifestyle interventions such as exercise and weight loss can slow OA progression, but at later stages, only an invasive option is available: total knee replacement (TKR). Though a generally successful procedure, only 2/3 of patients who undergo the procedure report their knees feeling “normal” post-operation, and complications can arise that require revision. This necessitates a model to identify a population at higher risk of TKR, particularly at less advanced stages of OA, such that appropriate treatments can be implemented that slow OA progression and delay TKR. Here, we present a deep learning pipeline that leverages MRI images and clinical and demographic information to predict TKR with AUC $0.834 \pm 0.036$ (p < 0.05). Most notably, the pipeline predicts TKR with AUC $0.943 \pm 0.057$ (p < 0.05) for patients without OA. Furthermore, we develop occlusion maps for case-control pairs in test data and compare regions used by the model in both, thereby identifying TKR imaging biomarkers. As such, this work takes strides towards a pipeline with clinical utility, and the biomarkers identified further our understanding of OA progression and eventual TKR onset.
\end{abstract}

\begin{document}

\flushbottom
\maketitle
%
%
\thispagestyle{empty}

\section*{Introduction}
\indent Knee Osteoarthritis (OA) is one of the most common musculoskeletal disorders in the United States, with estimates of its incidence rate ranging from 14 to 30 million \cite{Kremers2015, Deshpande2016}. Annual arthritis-related medical expenditures are nearly \$140 million, and hip and knee OA together are the 11th highest contributor to global disability \cite{Cross2014, Murphy2018}. The propensity of knee OA to induce eventual disability can be attributed to structural changes in the joint that characterize the disease, as well as symptoms that can include inflammation, debilitating pain, and functional limitations \cite{Lawrence2008, Ostrander2016}. Progression of the full-joint disease is typically assessed using the Kellgren-Lawrence (KL) scale, a 0-4 scale in which a higher score is associated with narrowing of the tibiofemoral joint (TFJ) space and other radiographic changes, and thus, a more advanced stage of knee OA \cite{Kellgren1957}. When diagnosed at early stages (KL = 0, 1), knee OA can be managed through nonsurgical treatment options, including exercise and/or weight loss, oral medications such as acetaminophen or NSAIDs, or intra-articular injections such as corticosteroids and hyaluronic acid, all of which have varying degrees of success in reducing pain \cite{Ringdahl2011}. At late stages (KL = 4), however, no noninvasive option exists \cite{Tiulpin2018}; here, the only option is total knee replacement (TKR).

TKR is an elective procedure in which the knee joint is resurfaced with a metal or plastic implant intended to restore function, provide pain relief, and improve quality of life \cite{Nguyen2015}. In the United States, estimates of TKR incidence lie at 400,000 each year, a figure expected to grow 143\% by 2050 even through conservative projections \cite{Inacio2017}. While TKR is considered one of the most effective procedures in orthopedic surgery, electing for it is far from straightforward: noninvasive alternatives such as weight loss, physical therapy, and NSAIDs are first exhausted. If unsuccessful, a patient will undergo a thorough examination of clinical history and comprehensive imaging of the joint to determine if a TKR is feasible, and if so, the desired implant design and size \cite{Tanzer2016, Sassoon2015}. The procedure is also imperfect: only 66\% of patients report their knees feeling “normal,” and 33\% of patients report some degree of pain post-implant \cite{Parvizi2014}. Furthermore, the implant can fail under some circumstances: periprosthetic joint infection and wound complications can be observed, and implant instability can occur due to aseptic loosening, malpositioning of the implant, and wear of joint components \cite{Chang2014, Kim2014}. It is thus much preferable to prolong the good health of the knee, particularly in patients where OA has not advanced to the most severe stages, thereby delaying TKR as long as possible. This necessitates a model to identify patients at higher risk of TKR such that appropriate treatment options can be pursued.

Given the multitude of factors on which a decision to pursue TKR is made, devising a model to predict if the invasive intervention will be necessary is a difficult task, but with obvious utility. For a patient in earlier stages of OA, a model predicting the patient to be at risk of TKR can be the impetus for a more aggressive nonsurgical treatment. Meanwhile, for a late-stage OA patient, a model predicting them to undergo TKR may facilitate a doctor and patient opting for the treatment earlier than they otherwise would, thereby reducing time spent pursuing nonsurgical alternatives with minimal probability of success while dealing with serious pain. Beyond this, if the model were to draw from medical images of the knee, it could identify anatomic regions most correlated with a TKR prediction. To this point, few studies have been conducted in this space, and those that have primarily investigate the importance of cartilage volume loss, subchondral bone defects, and bone marrow lesions \cite{Raynauld2008, Raynauld2013, Everhart2019}. An identification of more such biomarkers for TKR, however, could greatly improve understanding of both OA and TKR, and ultimately guide treatment strategies.

Predictive modeling of TKR, however, has a limited history, particularly with models that use medical images. A few studies have leveraged random forest regression, Cochran-Armitage tests for trend, and t-tests to identify demographic, general health, and physical examination measurements that most strongly correlate with TKR or total joint arthroplasty (TJA) \cite{Riddle2009, Hawker2006}. Others have taken these efforts further, using techniques such as multiple regression and multivariate risk prediction models to predict TKR outright \cite{Lewis2013, Yu2019}. To our knowledge, only one group has developed a predictive model of TKR that accepts image inputs, attaining performance that surpasses that of models using only clinical and demographic information \cite{Wang2019}. Notably, past TKR predictive models largely measure performance by evaluating the area under the receiver operating characteristic (ROC) curve, which plots true positive rate against false positive rate \cite{Akobeng2007}. However, in most datasets used in this space, the number of patients who eventually undergo TKR is dramatically higher among those who have advanced OA as opposed to those with no or moderate OA. Consequently, this performance metric (AUC), while effectively capturing a model’s combination of sensitivity and specificity, can be inflated for TKR prediction by indiscriminately predicting patients without OA not to undergo TKR, while more accurately predicting patients with severe OA to undergo TKR, the latter of which is easier. As a result, while past works have made clear progress in predicting TKR, none have overcome datasets imbalanced with respect to OA severity to report sensitive and specific prediction at these early stages, where a model would have the most utility.

One technique that has shown promise in delivering such performance is deep learning (DL). DL, especially convolutional neural networks (CNNs), has made strides in image classification tasks, attaining performances on the popular ImageNet classification challenge that approach or surpass human performance \cite{He2016, Huang2017, Russakovsky2015}. DL shines when afforded large datasets, as its automated feature extraction allows one to solve problems too complex for conventional approaches \cite{LeCun2015}. Given the complex prognostic features in TKR recommendation, CNNs become more promising for TKR prediction. In the past, DL had seen limited utility in OA and TKR prediction due to the large dataset requirement for efficacy; that limitation has been somewhat mitigated by the curation of large-sized cohort studies such as the Osteoarthritis Initiative (OAI) \cite{Peterfy2008}. Consequently, DL has recently been applied for knee OA classification and progression prediction \cite{Tiulpin2018, Antony2016, Norman2019}. The success of these works further suggests the feasibility of leveraging DL to predict TKR.

\begin{figure}[ht]
\centering
\includegraphics[width=\linewidth]{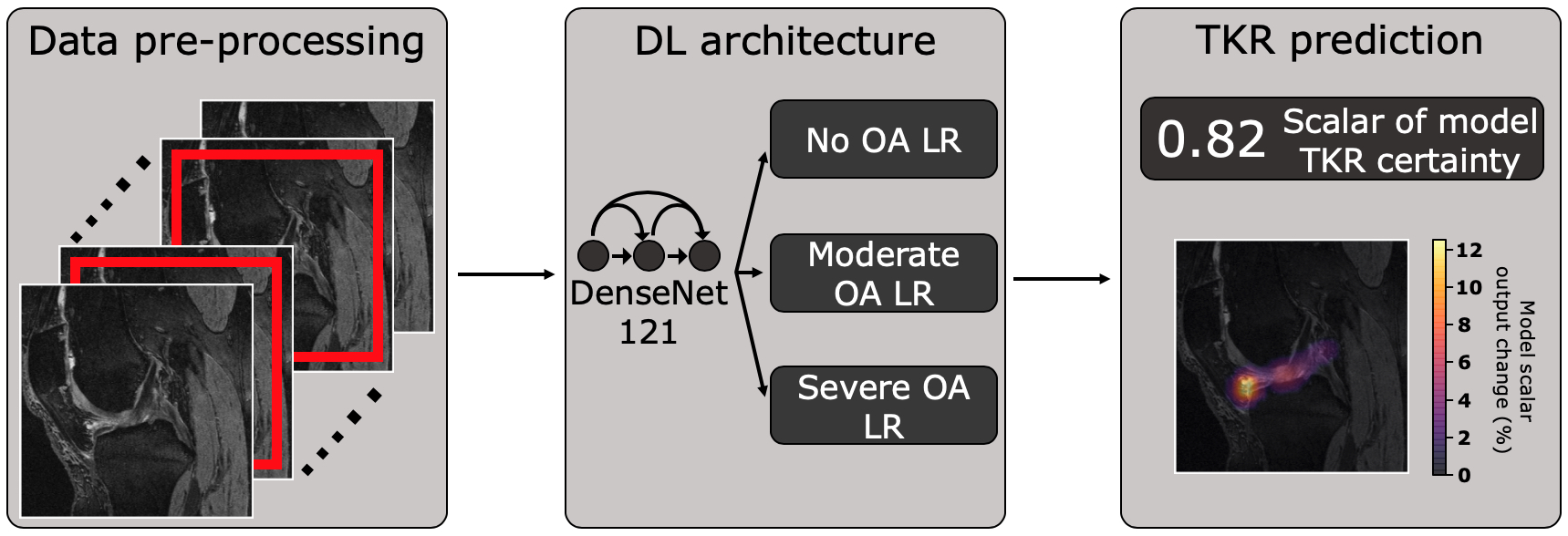}
\caption{Pipeline predicting if patient will undergo TKR within 5 years from MRI/X-ray images and non-imaging variables. MRI and X-ray images are center-cropped and cropped to a region centered around the joint, respectively, and normalized. DenseNet-121 is pretrained to predict OA and fine-tuned to predict TKR. Image-based predictions and clinical information are fed to a logistic regression (LR) ensemble based on OA severity. Each ensemble, whose hyperparameters were optimized for Youden’s index in a hyperparameter search, averages predictions of LR models in its OA severity for final TKR prediction. Pipeline is subsequently analyzed through occlusion map analysis to identify imaging biomarkers of TKR.}
\label{fig:pipeline_overview}
\end{figure}

In this study, we formulate a DL-based pipeline that incorporates knee joint images in addition to clinical and demographic information to predict the onset of TKR (Fig. \ref{fig:pipeline_overview}). We demonstrate that the pipeline’s predictions using solely Magnetic Resonance Imaging (MRI) images matches that of past work, while the integration of MRI image-based predictions with non-imaging variables facilitates TKR prediction with especially high sensitivity and specificity for patients without radiographic OA. Furthermore, we show the increase in pipeline performance when using 3D MRI images as opposed to 2D radiographs, suggesting MRI may have a role in TKR risk screening despite higher costs and more limited availability. And finally, we leverage occlusion maps to conduct a thorough analysis of tissues that most significantly affect the output model metric associated with TKR prediction confidence, thereby identifying a set of imaging biomarkers for eventual TKR onset.

\section*{Novelty}
This work reports a methodology and results that are novel in the following manners: 
\begin{enumerate}
\item This model is the first to apply a 3-dimensional DenseNet CNN for prediction of TKR from MRI.
\item The TKR prediction model is evaluated for patients stratified by OA severity, which has not been reported in previous studies. 
\item With the aim of improving model interpretability and clinical utility, we report the first comprehensive, case-control study to identify imaging biomarkers for TKR.
\end{enumerate}

\section*{Materials and Methods}
\subsection*{Data}
Data was acquired from a prospective observational study conducted by OAI. The dataset followed 4,796 patients and acquired images including 2D posteroanterior radiographs and 3D Sagittal Double Echo Steady-State (DESS) MRI images over the course of 10 years. Details of data collection and study design have been previously reported \cite{Peterfy2008}. The OAI study protocol was approved by the National Institute of Arthritis and Musculoskeletal and Skin Diseases (NIAMS) and is registered on ClinicalTrials.gov as “Osteoarthritis Initiative (OAI): A Knee Health Study”, NCT\#00080171. The study was carried out in accordance with all pertinent guidelines, and written consent was obtained from participants prior to each clinical visit in the study. 

Both posteroanterior radiographs and DESS MRI images were evaluated as data sources for TKR prediction models. Patients for whom KL grade was not recorded at any point in the longitudinal study were excluded. To homogenize datasets, radiograph and MRI images were only taken from patients and time points at which both were available (n = 35,482). We labeled entries as cases if the patient underwent a first TKR within 5 years of the given time point (n = 1,043). We labeled entries as controls if patients did not undergo a TKR or eventually underwent one but the time to it was longer than 5 years (n = 34,439). Contralateral TKRs were not considered.

The radiographs and MRI images were preprocessed for training and model evaluation. Radiographs were cropped to a $500 \times 500$ region centered around the knee joint. Briefly, 2D cross-correlation template matching was used to identify a $500 \times 500$ bounding box centered around the knee joint in 450 joints, and these cases were used to train a U-Net architecture that identified this region for all posteroanterior radiographs from the OAI study \cite{Norman2019}.  DESS MRIs were center-cropped to a $120 \times 320 \times 320$ region, after which both sets of cropped images were normalized. Normalized MRI pixel values were then rounded to nearest integers, compressing the MRI image to 14 possible pixel values. This rounding approach was initially tested as a strategy to accelerate training of a 3D CNN, given the large imaging volumes and large dataset on which it was being trained, believing the approach could suppress information extraneous to eventual TKR. Empirically, this approach yielded superior validation performance to leaving pixel values unrounded, so it was utilized. Examples of the results of this compression strategy are in Supplementary Fig. \ref{fig:image_compression}. 

Non-imaging variables were screened for among studies and reviews detailing risk factors for knee OA progression and TKR onset \cite{Hawker2006, Lewis2013, Yu2019, Heidari2011, Cooper2000, Pisters2012, Sharma2003}. Variables such as KL grade known to be deducible directly from MRI images and radiographs were not considered. From these studies, 40 non-imaging variables of interest were identified (see Supplementary Table \ref{tab:potential_variable_list}). The OAI database was then parsed for corresponding variables, and these corresponding variables were added as potential non-imaging variables for our study, yielding 44 potential non-imaging variables. In some cases, multiple OAI metrics corresponded to non-imaging variables of interest, causing the number of OAI non-imaging variables to exceed what was identified from literature. Missing data points were imputed with k-nearest neighbors. These potential variables were used to train a random forest with 100 trees to predict onset of TKR within 5 years, and the minimum depth at which each feature was used across all trees in the forest was identified. Features whose minimum depth was below the average minimum depth of all features were preserved as non-imaging variables \cite{Dietrich2016}. This yielded 27 non-imaging variables that are displayed in Table \ref{tab:non_imaging_vars}.

  \begin{table}[t]
\centering
\begin{tabular}{ |l l| } 
\hline
\multicolumn{2}{|c|}{\textbf{Non-imaging variables used to augment image-based predictions}}\\
\hline
Age & Comorbidity score\\
BMI & Injections to treat arthritis in previous 6 months\\
Education & Seen physician for arthritis in previous year\\
Ethnicity & Knee valgus negative alignment (degrees)\\
Income & Isometric leg strength\\
NSAID usage & Back pain in previous 30 days\\
Analgesics usage & Difficulty squatting in previous 7 days\\
Systolic BP & Difficulty kneeling in previous 7 days\\
Considering TKR & Baseline frequency knee pain status\\
PASE & Previous knee injury that limited walking\\
KOOS QOL & 0-10 global rating assessing effect of knee pain\\
KOOS pain & SF-12 physical component score\\
WOMAC pain & SF-12 mental component score\\
WOMAC disability & \\
\hline
\end{tabular}
\caption{\label{tab:non_imaging_vars}List of non-imaging variables fed into logistic regression models to make TKR onset predictions. Abbreviations used: Body Mass Index (BMI), Nonsteroidal Anti-inflammatory drugs (NSAIDS), Blood Pressure (BP), Physical Activity Scale for the Elderly (PASE), Knee Injury and Osteoarthritis Outcome Score (KOOS), Quality of Life (QOL), Western Ontario and McMaster Universities Arthritis Index (WOMAC), Short Form 12 (SF-12).}
\end{table}

The data were then split into training, validation, and test with a 65\%/20\%/15\% split, ensuring entries of any patient were only in one of the three datasets to prevent data leakage. Within the training set, imbalance between TKR and non-TKR cases was addressed with data augmentation, drawing bootstrap samples from the rare class with replacement \cite{Chen2004}. A summary of the data prior to augmentation is provided in Table \ref{tab:data_description}, detailing the number of cases and controls while showing descriptive statistics regarding demographics in each of the three datasets.

\begin{table}[b!]
\centering
\begin{tabular}{ |P{0.7cm}|P{0.3cm}|P{0.85cm}|P{0.85cm}|P{0.85cm}|P{0.7cm}|P{0.7cm}|P{0.6cm}|P{0.8cm}|P{0.8cm}|P{0.7cm}|P{0.8cm}| } 
\hline
\multicolumn{1}{|c|}{\multirow{2}{*}{\textbf{Dataset}}}&\multicolumn{1}{c|}{\multirow{1}[4]{*}{\begin{tabular}{@{}c@{}}\textbf{Data} \\ \textbf{type}\end{tabular}}}&\multicolumn{1}{c|}{\multirow{2}{*}{\textbf{Age}}}&\multicolumn{1}{c|}{\multirow{2}{*}{\textbf{BMI}}}&\multicolumn{1}{c|}{\multirow{1}[4]{*}{\begin{tabular}{@{}c@{}}\textbf{KOOS} \\ \textbf{pain}\end{tabular}}}&\multicolumn{1}{c|}{\multirow{2}{*}{\textbf{Male}}}&\multicolumn{1}{c|}{\multirow{2}{*}{\textbf{Female}}}&\multicolumn{3}{c|}{\textbf{OA status}}&\multicolumn{1}{c|}{\multirow{1}[4]{*}{\begin{tabular}{@{}c@{}}\textbf{Total} \\ \textbf{entries}\end{tabular}}}&\multicolumn{1}{c|}{\multirow{1}[4]{*}{\begin{tabular}{@{}c@{}}\textbf{Unique} \\ \textbf{patients}\end{tabular}}}\\
\cline{8-10}
&&&&&&&\multicolumn{1}{c|}{\textbf{None}}&\multicolumn{1}{c|}{\textbf{Moderate}}&\multicolumn{1}{c|}{\textbf{Severe}}&&\\
\hline
\multicolumn{1}{|c|}{\multirow{2}[8]{*}{Training}}&\multicolumn{1}{c|}{\multirow{1}[4]{*}{Control}}&$62.5\pm9.15$&\multicolumn{1}{c|}{\multirow{1}[4]{*}{\begin{tabular}{@{}c@{}}$28.3\pm$ \\ 4.75\end{tabular}}}&\multicolumn{1}{c|}{\multirow{1}[4]{*}{\begin{tabular}{@{}c@{}}$87.2\pm$ \\ 16.2\end{tabular}}}&\multicolumn{1}{c|}{\multirow{1}[4]{*}{9,708}}&\multicolumn{1}{c|}{\multirow{1}[4]{*}{12,731}}&\multicolumn{1}{c|}{\multirow{1}[4]{*}{12,721}}&\multicolumn{1}{c|}{\multirow{1}[4]{*}{8,950}}&\multicolumn{1}{c|}{\multirow{1}[4]{*}{768}}&\multicolumn{1}{c|}{\multirow{2}[8]{*}{23,126}}&\multicolumn{1}{c|}{\multirow{2}[8]{*}{3,114}}\\
&\multicolumn{1}{c|}{\multirow{1}[4]{*}{Case}}&$66.3\pm8.38$&\multicolumn{1}{c|}{\multirow{1}[4]{*}{\begin{tabular}{@{}c@{}}$29.6\pm$ \\ 4.79\end{tabular}}}&\multicolumn{1}{c|}{\multirow{1}[4]{*}{\begin{tabular}{@{}c@{}}$67.2\pm$ \\ 19.6\end{tabular}}}&\multicolumn{1}{c|}{\multirow{1}[4]{*}{291}}&\multicolumn{1}{c|}{\multirow{1}[4]{*}{396}}&\multicolumn{1}{c|}{\multirow{1}[4]{*}{41}}&\multicolumn{1}{c|}{\multirow{1}[4]{*}{357}}&\multicolumn{1}{c|}{\multirow{1}[4]{*}{289}}&&\\
\hline
\multicolumn{1}{|c|}{\multirow{2}[8]{*}{Valid.}}&\multicolumn{1}{c|}{\multirow{1}[4]{*}{Control}}&$62.4\pm9.21$&\multicolumn{1}{c|}{\multirow{1}[4]{*}{\begin{tabular}{@{}c@{}}$28.4\pm$ \\ 4.64\end{tabular}}}&\multicolumn{1}{c|}{\multirow{1}[4]{*}{\begin{tabular}{@{}c@{}}$87.6\pm$ \\ 15.9\end{tabular}}}&\multicolumn{1}{c|}{\multirow{1}[4]{*}{2,876}}&\multicolumn{1}{c|}{\multirow{1}[4]{*}{4,035}}&\multicolumn{1}{c|}{\multirow{1}[4]{*}{4,118}}&\multicolumn{1}{c|}{\multirow{1}[4]{*}{2,611}}&\multicolumn{1}{c|}{\multirow{1}[4]{*}{182}}&\multicolumn{1}{c|}{\multirow{2}[8]{*}{7,115}}&\multicolumn{1}{c|}{\multirow{2}[8]{*}{957}}\\
&\multicolumn{1}{c|}{\multirow{1}[4]{*}{Case}}&$66.1\pm8.76$&\multicolumn{1}{c|}{\multirow{1}[4]{*}{\begin{tabular}{@{}c@{}}$29.8\pm$ \\ 4.61\end{tabular}}}&\multicolumn{1}{c|}{\multirow{1}[4]{*}{\begin{tabular}{@{}c@{}}$66.2\pm$ \\ 19.1\end{tabular}}}&\multicolumn{1}{c|}{\multirow{1}[4]{*}{70}}&\multicolumn{1}{c|}{\multirow{1}[4]{*}{134}}&\multicolumn{1}{c|}{\multirow{1}[4]{*}{13}}&\multicolumn{1}{c|}{\multirow{1}[4]{*}{93}}&\multicolumn{1}{c|}{\multirow{1}[4]{*}{98}}&&\\
\hline
\multicolumn{1}{|c|}{\multirow{2}[8]{*}{Test}}&\multicolumn{1}{c|}{\multirow{1}[4]{*}{Control}}&$62.8\pm9.55$&\multicolumn{1}{c|}{\multirow{1}[4]{*}{\begin{tabular}{@{}c@{}}$28.4\pm$ \\ 4.81\end{tabular}}}&\multicolumn{1}{c|}{\multirow{1}[4]{*}{\begin{tabular}{@{}c@{}}$87.4\pm$ \\ 16.5\end{tabular}}}&\multicolumn{1}{c|}{\multirow{1}[4]{*}{2,126}}&\multicolumn{1}{c|}{\multirow{1}[4]{*}{2,963}}&\multicolumn{1}{c|}{\multirow{1}[4]{*}{2,892}}&\multicolumn{1}{c|}{\multirow{1}[4]{*}{2,056}}&\multicolumn{1}{c|}{\multirow{1}[4]{*}{141}}&\multicolumn{1}{c|}{\multirow{2}[8]{*}{5,241}}&\multicolumn{1}{c|}{\multirow{2}[8]{*}{719}}\\
&\multicolumn{1}{c|}{\multirow{1}[4]{*}{Case}}&$66.4\pm7.78$&\multicolumn{1}{c|}{\multirow{1}[4]{*}{\begin{tabular}{@{}c@{}}$29.9\pm$ \\ 3.96\end{tabular}}}&\multicolumn{1}{c|}{\multirow{1}[4]{*}{\begin{tabular}{@{}c@{}}$68.7\pm$ \\ 20.6\end{tabular}}}&\multicolumn{1}{c|}{\multirow{1}[4]{*}{59}}&\multicolumn{1}{c|}{\multirow{1}[4]{*}{93}}&\multicolumn{1}{c|}{\multirow{1}[4]{*}{12}}&\multicolumn{1}{c|}{\multirow{1}[4]{*}{83}}&\multicolumn{1}{c|}{\multirow{1}[4]{*}{57}}&&\\
\hline
\end{tabular}
\caption{\label{tab:data_description}Data used to train 3D DESS MRI and 2D radiograph architectures. After exclusion criteria were applied, 35,482 qualifying entries were found in the OAI dataset across 4,790 unique patients, all of which were split into training, validation, and test sets as displayed in table. To prevent data leakage, all entries from any given patient were only allowed to be in one of the three sets. S.d. is reported for age, BMI, and KOOS pain score within the table.}
\end{table}

\subsection*{Pipeline architecture}
The DL-based pipeline is based on a DenseNet-121 with the following parameters: 16 filters in initial layer, growth rate of 32, pooling block configuration of [6, 12, 24, 16], 4 bottleneck layers, 2 classes. The same architecture was used for the radiograph and MRI pipelines, but for the MRI pipeline, we modified the convolutional layers, batch normalization layers, pooling layers, and leaky rectified linear unit (ReLU) layers to allow for 3D image input \cite{Hara2018}. The network yielded a scalar reflecting certainty of TKR within 5 years, which was added to the non-imaging variables. The 28 resulting variables were fed into one of three sets of Logistic Regression (LR) ensembles, with each ensemble optimized to maximize sensitivity and specificity in cases of no (KL = 0, 1), moderate (KL = 2, 3), and severe OA (KL = 4). Based on the KL grade of a sample, it was fed into an LR ensemble, yielding a prediction as to whether the patient will undergo a TKR within 5 years.

\subsection*{Training}
A DenseNet-121 was initially pretrained to predict knee OA using the entire training set, assessing cross-entropy loss and accuracy on the validation set after completion of each epoch. The pre-train was stopped when validation loss began to increase. The pretrained model was subsequently fine-tuned to predict TKR. We utilized a random search to determine optimal learning rate, dropout rate, weights of the cross-entropy loss function, and number of layers to freeze during fine-tuning. The search was carried out for 25 iterations, after which a set of parameters were selected that yielded the best combination of accuracy, sensitivity, and specificity on the validation set. Due to computational intensity, the hyperparameter search was not conducted on the entire dataset: for the 2D DenseNet-121, 10\% of training and validation sets were used, whereas for the 3D DenseNet-121, 2.5\% of both were used. After the search, the model fine-tuned using the subset of the training set was further fine-tuned on the entire training set using optimal parameters until validation loss began to increase. The test set was held out during training and predictions for it evaluated just once after fine-tuning, which marked the end of model optimization.

\subsection*{Integration of Imaging and Non-Imaging Data}
Random forest regression, support vector machine, neural network, and LR architectures were assessed for efficacy of integrating imaging and non-imaging predictions, with LR providing best results on validation data. The LR architecture was thus used: all 28 imaging and non-imaging models were fed into an LR model, the optimal parameters of which were also identified through a random search. The search was conducted for 100 iterations, seeking to optimize the cross-entropy loss function weights afforded to both classes. For the cases of no, moderate, and severe OA, ideal parameters were identified by selecting those that maximized Youden’s index within each OA classification in the search \cite{Youden1950}. Predictions of the best few models in each classification were averaged to yield final TKR predictions. The number of predictions averaged in each classification was selected by finding a value that optimized validation accuracy, AUC, and Youden’s index. The resulting LR models were ensembled and run on test data just once. Confidence intervals of accuracy, sensitivity, and specificity for each OA severity were obtained by bootstrapping, sampling 100\% of test data with replacement (B = 100). Confidence intervals for AUC were calculated in the same manner. Results are reported on 3 versions of each model: the sole DenseNet-121 output (image only), output of a single LR model trained to predict TKR using solely the 27 non-imaging variables while not weighting the loss function class weights (non-imaging info. only), and output of the LR ensemble with image predictions (integrated model).

\subsection*{Statistical Analysis}
The accuracies of X-ray and MRI pipeline performances within each OA classification and overall were compared using McNemar’s test \cite{Salzberg1997, Dietterich1998}. This test was appropriate because it specifically tests for differences in a dichotomous variable in matched groups. In our case, the variable was correct TKR prediction and the groups were the X-ray and MRI pipelines. Initially, the McNemar test statistic was modeled with a chi-squared distribution to test for significant differences between the pipelines, and if one existed, a binomial distribution was used to interrogate which pipeline yielded the significantly higher performance. All tests were carried out at $\alpha = 0.05$. 

Relative sensitivity and specificity of the X-ray and MRI pipelines were assessed by comparing their AUCs within each OA classification and overall. This test is appropriate because the ROC curve plots true positive rate (sensitivity) against false positive rate (1 – specificity); consequently, the closer the AUC is to 1, the better the combination of sensitivity and specificity. 100\% of test data was sampled with replacement (B = 100), and for each corresponding pair of X-ray and MRI pipelines (matched by OA classification and use of images only or both image and non-image information), AUCs were calculated. To test if one outperformed the other, differences in AUCs were calculated at each iteration, and the mean and standard deviation of the differences used to conduct a student’s t-test with 99 degrees of freedom. This test is applicable on each matched pair of X-ray and MRI pipelines due to the number of iterations for which test data was sampled, allowing the central limit theorem to apply. For confidence intervals, mean and standard deviation of AUCs of individual models were calculated and used to report 95\% intervals.
 
\subsection*{Imaging biomarker identification}
For all 124 true positives in the test data for the integrated MRI pipeline, corresponding controls were identified by randomly sampling from test data true negatives, keeping OA status distributions identical and using a student’s t-test with 123 degrees of freedom to ensure no significant difference in KOOS pain scores across cases and corresponding controls at $\alpha = 0.05$. Occlusion maps were generated for all cases and controls using voxel size of $12 \times 32 \times 32$ and stride of 12. For each pixel, the value displayed represented the magnitude of change in the scalar pipeline output resulting when that pixel was occluded, averaged across all occlusions in which that pixel existed. Pixels for which scalar pipeline output change lied in the top 5\% were designated as “hotspots.” Anatomic regions of these hotspots were identified and odds ratios (OR) calculated to interrogate possible imaging biomarkers of TKR. 95\% OR confidence intervals were calculated for each anatomic region investigated in this analysis using Cornfield’s method, as this method performs well with relatively small sample sizes \cite{Lawson2004}. P values of ORs were calculated using a two-tailed Fisher’s exact test \cite{Upton1992}. Tissues where p values fell below the significance level of $\alpha = 0.05$ and in which 95\% OR confidence intervals did not include 1 were deemed significant. These test selections were appropriate, as they allowed for direct comparison of the frequencies at which several tissues were hotspots across cases and controls, and as such, identified significant tissues with regards to TKR onset.

\section*{Results}
\subsection*{OA pretrain utility in TKR prediction}
To test information learned from the OA pretrain, pretrained models themselves were used to predict TKR, with results depicted in Table \ref{tab:oa_pretrain_perf}. Predictably, the radiograph OA pretrain model had poor sensitivity for patients without OA, and poor specificity in moderate and severe cases of OA. While the MRI OA pretrain model expectedly yielded more balanced sensitivity and specificity across all OA stages, it too left room for improvement, particularly in sensitivity at no OA and specificity at severe OA. This confirmed the pretrain provided useful information to both architectures but fine-tuning and integration of non-imaging variables were necessary to attain desired TKR prediction performance. 

\begin{table}[t!]
\centering
\begin{tabular}{ |P{0.7cm}|P{0.3cm}|P{1.9cm}|P{1.9cm}|P{1.9cm}|P{0.7cm}|P{0.7cm}| } 
\hline
\multicolumn{1}{|c|}{\multirow{1}{*}{\begin{tabular}{@{}c@{}}\textbf{OA} \\ \textbf{status}\end{tabular}}}&\multicolumn{1}{c|}{\multirow{1}{*}{\begin{tabular}{@{}c@{}}\textbf{Model} \\ \textbf{type}\end{tabular}}}&\multicolumn{1}{c|}{\multirow{1}{*}{\begin{tabular}{@{}c@{}}\textbf{Accuracy} \\ \textbf{(95\% CI)}\end{tabular}}}&\multicolumn{1}{c|}{\multirow{1}{*}{\begin{tabular}{@{}c@{}}\textbf{Sensitivity} \\ \textbf{(95\% CI)}\end{tabular}}}&\multicolumn{1}{c|}{\multirow{1}{*}{\begin{tabular}{@{}c@{}}\textbf{Specificity} \\ \textbf{(95\% CI)}\end{tabular}}}&\multicolumn{1}{c|}{\multirow{1}{*}{\begin{tabular}{@{}c@{}}\textbf{Non-TKR} \\ \textbf{cases}\end{tabular}}}&\multicolumn{1}{c|}{\multirow{1}{*}{\begin{tabular}{@{}c@{}}\textbf{TKR} \\ \textbf{cases}\end{tabular}}}\\
&&&&&&\\
\hline
\multicolumn{1}{|c|}{\multirow{2}[0]{*}{None}}&\multicolumn{1}{c|}{\multirow{1}[1]{*}{Radiograph}}&\multicolumn{1}{c|}{\multirow{1}[1]{*}{$92.1\pm0.083$}}&\multicolumn{1}{c|}{\multirow{1}[1]{*}{$25.2\pm2.16$}}&\multicolumn{1}{c|}{\multirow{1}[1]{*}{$92.4\pm0.081$}}&\multicolumn{1}{c|}{\multirow{2}[0]{*}{2,892}}&\multicolumn{1}{c|}{\multirow{2}[0]{*}{12}}\\
&\multicolumn{1}{c|}{\multirow{2}[-3]{*}{MRI}}&\multicolumn{1}{c|}{\multirow{1}[1]{*}{$94.3\pm0.070$}}&\multicolumn{1}{c|}{\multirow{1}[1]{*}{$48.7\pm2.48$}}&\multicolumn{1}{c|}{\multirow{1}[1]{*}{$94.4\pm0.070$}}&&\\
\hline
\multicolumn{1}{|c|}{\multirow{2}[0]{*}{Moderate}}&\multicolumn{1}{c|}{\multirow{1}[1]{*}{Radiograph}}&\multicolumn{1}{c|}{\multirow{1}[1]{*}{$29.3\pm0.151$}}&\multicolumn{1}{c|}{\multirow{1}[1]{*}{$93.8\pm0.439$}}&\multicolumn{1}{c|}{\multirow{1}[1]{*}{$26.7\pm0.156$}}&\multicolumn{1}{c|}{\multirow{2}[0]{*}{2,056}}&\multicolumn{1}{c|}{\multirow{2}[0]{*}{83}}\\
&\multicolumn{1}{c|}{\multirow{2}[-3]{*}{MRI}}&\multicolumn{1}{c|}{\multirow{1}[1]{*}{$65.4\pm0.154$}}&\multicolumn{1}{c|}{\multirow{1}[1]{*}{$65.5\pm0.848$}}&\multicolumn{1}{c|}{\multirow{1}[1]{*}{$65.4\pm0.158$}}&&\\
\hline
\multicolumn{1}{|c|}{\multirow{2}[0]{*}{Severe}}&\multicolumn{1}{c|}{\multirow{1}[1]{*}{Radiograph}}&\multicolumn{1}{c|}{\multirow{1}[1]{*}{$29.7\pm0.488$}}&\multicolumn{1}{c|}{\multirow{1}[1]{*}{$100.0\pm0.000$}}&\multicolumn{1}{c|}{\multirow{1}[1]{*}{$1.4\pm0.180$}}&\multicolumn{1}{c|}{\multirow{2}[0]{*}{141}}&\multicolumn{1}{c|}{\multirow{2}[0]{*}{57}}\\
&\multicolumn{1}{c|}{\multirow{2}[-3]{*}{MRI}}&\multicolumn{1}{c|}{\multirow{1}[1]{*}{$33.4\pm0.523$}}&\multicolumn{1}{c|}{\multirow{1}[1]{*}{$82.2\pm0.824$}}&\multicolumn{1}{c|}{\multirow{1}[1]{*}{$14.0\pm0.441$}}&&\\
\hline
\multicolumn{1}{|c|}{\multirow{2}[0]{*}{All}}&\multicolumn{1}{c|}{\multirow{1}[1]{*}{Radiograph}}&\multicolumn{1}{c|}{\multirow{1}[1]{*}{$64.2\pm0.124$}}&\multicolumn{1}{c|}{\multirow{1}[1]{*}{$90.7\pm0.378$}}&\multicolumn{1}{c|}{\multirow{1}[1]{*}{$63.4\pm0.126$}}&\multicolumn{1}{c|}{\multirow{2}[0]{*}{5,089}}&\multicolumn{1}{c|}{\multirow{2}[0]{*}{152}}\\
&\multicolumn{1}{c|}{\multirow{2}[-3]{*}{MRI}}&\multicolumn{1}{c|}{\multirow{1}[1]{*}{$80.2\pm0.079$}}&\multicolumn{1}{c|}{\multirow{1}[1]{*}{$70.4\pm0.595$}}&\multicolumn{1}{c|}{\multirow{1}[1]{*}{$80.5\pm0.082$}}&&\\
\hline
\end{tabular}
\caption{\label{tab:oa_pretrain_perf}Performance in TKR prediction of OA pretrained models for radiographs and MRI, stratified by severity of OA. Pretraining strategy yields useful information to both models, but performance at no OA in particular leaves room for improvement, justifying subsequent model fine-tuning. Standard errors used to calculate confidence intervals.}
\end{table}

\subsection*{X-ray pipeline optimization and performance}
For the X-Ray model, hyperparameter tuning steps found the following to yield the best combination of validation accuracy, sensitivity, and specificity: learning rate of \num{3.981e-6}, TKR class weight in cross-entropy loss function of 0.927 and non-TKR class weight of 0.073, dropout rate of 0.375, and only the last 2 layers fine-tuned after OA pretrain.

\begin{figure}[ht]
\centering
\includegraphics[width=\linewidth]{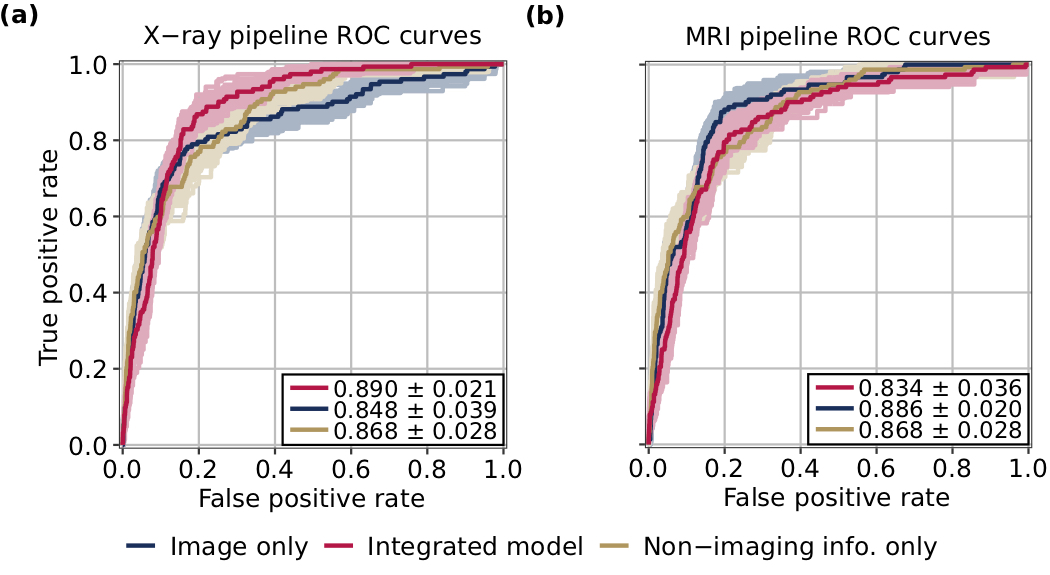}
\caption{ROC curves for X-ray and MRI architectures on test data. X-ray pipeline ROC curves are shown in \textbf{(a)}, with AUCs as follows, p < 0.05: $0.848 \pm 0.039$ (image only), $0.868 \pm 0.028$ (non-imaging info. only), $0.890 \pm 0.021$ (integrated model). MRI pipeline ROC curves are shown in \textbf{(b)}, with AUCs as follows, p < 0.05: $0.886 \pm 0.020$ (image only), $0.868 \pm 0.028$ (non-imaging info. only), $0.834 \pm 0.036$ (integrated model). Standard deviations used to calculate confidence intervals. ROC curves with AUCs within 1 standard deviation of the mean for each model type during bootstrapping are also shown on plots.}
\label{fig:xray_mri_performance}
\end{figure}

A radiograph model was fine-tuned to predict TKR with these parameters, and its predictions fed into an LR ensemble. Averaging predictions of the best 5 LR models found through random search in the 3 OA categories yielded best validation performance, so this ensemble was used on the test set. Test accuracy, sensitivity, and specificity are provided in Table \ref{tab:model_perf}, and ROC curves of all three versions of this pipeline are found in Fig. \ref{fig:xray_mri_performance}a. AUCs are as follows: $0.848 \pm 0.039$ (image only), $0.868 \pm 0.028$ (non-imaging info. only), $0.890 \pm 0.021$ (integrated model). Furthermore, AUCs for the image-only and combined versions of the pipeline at no OA are as follows: $0.514 \pm 0.087$ (image only); $0.799 \pm 0.055$ (integrated model). At moderate OA: $0.788 \pm 0.025$ (image only); $0.865 \pm 0.016$ (integrated model). At severe OA: $0.552 \pm 0.040$ (image only); $0.641 \pm 0.044$ (integrated model). All AUC intervals are calculated using standard deviation (s.d.), p < 0.05.

\begin{table}[t]
\centering
\begin{tabular}{ |P{0.7cm}|P{0.3cm}|P{0.8cm}|P{1.9cm}|P{1.9cm}|P{1.9cm}|P{0.7cm}|P{0.7cm}|} 
\hline
\multicolumn{1}{|c|}{\multirow{1}{*}{\begin{tabular}{@{}c@{}}\textbf{OA} \\ \textbf{status}\end{tabular}}}&\multicolumn{1}{c|}{\multirow{1}{*}{\begin{tabular}{@{}c@{}}\textbf{Image} \\ \textbf{source}\end{tabular}}}&\multicolumn{1}{c|}{\multirow{1}[4]{*}{\textbf{Model type}}}&\multicolumn{1}{c|}{\multirow{1}{*}{\begin{tabular}{@{}c@{}}\textbf{Accuracy} \\ \textbf{(95\% CI)}\end{tabular}}}&\multicolumn{1}{c|}{\multirow{1}{*}{\begin{tabular}{@{}c@{}}\textbf{Sensitivity} \\ \textbf{(95\% CI)}\end{tabular}}}&\multicolumn{1}{c|}{\multirow{1}{*}{\begin{tabular}{@{}c@{}}\textbf{Specificity} \\ \textbf{(95\% CI)}\end{tabular}}}&\multicolumn{1}{c|}{\multirow{1}{*}{\begin{tabular}{@{}c@{}}\textbf{Non-TKR} \\ \textbf{cases}\end{tabular}}}&\multicolumn{1}{c|}{\multirow{1}{*}{\begin{tabular}{@{}c@{}}\textbf{TKR} \\ \textbf{cases}\end{tabular}}}\\
&&&&&&&\\
\hline
\multicolumn{1}{|c|}{\multirow{6}[0]{*}{None}}&\multicolumn{1}{c|}{\multirow{3}[0]{*}{X-ray}}&\multicolumn{1}{c|}{\multirow{1}[0]{*}{Non-imaging info. only}}&\multicolumn{1}{c|}{\multirow{1}[0]{*}{$89.1\pm0.139$}}&\multicolumn{1}{c|}{\multirow{1}[0]{*}{$49.7\pm2.80$}}&\multicolumn{1}{c|}{\multirow{1}[0]{*}{$89.3\pm0.140$}}&\multicolumn{1}{c|}{\multirow{6}[0]{*}{2,892}}&\multicolumn{1}{c|}{\multirow{6}[0]{*}{12}}\\
&&\multicolumn{1}{c|}{\multirow{1}[0]{*}{Image only}}&\multicolumn{1}{c|}{\multirow{1}[0]{*}{$95.0\pm0.089$}}&\multicolumn{1}{c|}{\multirow{1}[0]{*}{$7.8\pm1.64$}}&\multicolumn{1}{c|}{\multirow{1}[0]{*}{$95.4\pm0.089$}}&&\\
&&\multicolumn{1}{c|}{\multirow{1}[0]{*}{Integrated Model}}&\multicolumn{1}{c|}{\multirow{1}[0]{*}{$95.4\pm0.081$}}&\multicolumn{1}{c|}{\multirow{1}[0]{*}{$8.6\pm1.95$}}&\multicolumn{1}{c|}{\multirow{1}[0]{*}{$95.8\pm0.077$}}&&\\
\cline{2-6}
&\multicolumn{1}{c|}{\multirow{3}[0]{*}{MRI}}&\multicolumn{1}{c|}{\multirow{1}[0]{*}{Non-imaging info. only}}&\multicolumn{1}{c|}{\multirow{1}[0]{*}{$89.1\pm0.139$}}&\multicolumn{1}{c|}{\multirow{1}[0]{*}{$49.7\pm2.80$}}&\multicolumn{1}{c|}{\multirow{1}[0]{*}{$89.3\pm0.140$}}&&\\
&&\multicolumn{1}{c|}{\multirow{1}[0]{*}{Image only}}&\multicolumn{1}{c|}{\multirow{1}[0]{*}{$95.2\pm0.088$}}&\multicolumn{1}{c|}{\multirow{1}[0]{*}{$66.9\pm3.23$}}&\multicolumn{1}{c|}{\multirow{1}[0]{*}{$95.3\pm0.089$}}&&\\
&&\multicolumn{1}{c|}{\multirow{1}[0]{*}{Integrated Model}}&\multicolumn{1}{c|}{\multirow{1}[0]{*}{$82.4\pm0.171$}}&\multicolumn{1}{c|}{\multirow{1}[0]{*}{$92.2\pm1.68$}}&\multicolumn{1}{c|}{\multirow{1}[0]{*}{$82.4\pm0.173$}}&&\\
\hline
\multicolumn{1}{|c|}{\multirow{6}[0]{*}{Moderate}}&\multicolumn{1}{c|}{\multirow{3}[0]{*}{X-ray}}&\multicolumn{1}{c|}{\multirow{1}[0]{*}{Non-imaging info. only}}&\multicolumn{1}{c|}{\multirow{1}[0]{*}{$72.9\pm0.208$}}&\multicolumn{1}{c|}{\multirow{1}[0]{*}{$70.0\pm1.16$}}&\multicolumn{1}{c|}{\multirow{1}[0]{*}{$73.0\pm0.212$}}&\multicolumn{1}{c|}{\multirow{6}[0]{*}{2,056}}&\multicolumn{1}{c|}{\multirow{6}[0]{*}{83}}\\
&&\multicolumn{1}{c|}{\multirow{1}[0]{*}{Image only}}&\multicolumn{1}{c|}{\multirow{1}[0]{*}{$79.9\pm0.196$}}&\multicolumn{1}{c|}{\multirow{1}[0]{*}{$66.7\pm1.23$}}&\multicolumn{1}{c|}{\multirow{1}[0]{*}{$80.4\pm0.195$}}&&\\
&&\multicolumn{1}{c|}{\multirow{1}[0]{*}{Integrated Model}}&\multicolumn{1}{c|}{\multirow{1}[0]{*}{$81.4\pm0.178$}}&\multicolumn{1}{c|}{\multirow{1}[0]{*}{$76.0\pm1.12$}}&\multicolumn{1}{c|}{\multirow{1}[0]{*}{$81.6\pm0.179$}}&&\\
\cline{2-6}
&\multicolumn{1}{c|}{\multirow{3}[0]{*}{MRI}}&\multicolumn{1}{c|}{\multirow{1}[0]{*}{Non-imaging info. only}}&\multicolumn{1}{c|}{\multirow{1}[0]{*}{$72.9\pm0.208$}}&\multicolumn{1}{c|}{\multirow{1}[0]{*}{$70.0\pm1.16$}}&\multicolumn{1}{c|}{\multirow{1}[0]{*}{$73.0\pm0.212$}}&&\\
&&\multicolumn{1}{c|}{\multirow{1}[0]{*}{Image only}}&\multicolumn{1}{c|}{\multirow{1}[0]{*}{$68.8\pm0.225$}}&\multicolumn{1}{c|}{\multirow{1}[0]{*}{$78.3\pm0.952$}}&\multicolumn{1}{c|}{\multirow{1}[0]{*}{$68.4\pm0.227$}}&&\\
&&\multicolumn{1}{c|}{\multirow{1}[0]{*}{Integrated Model}}&\multicolumn{1}{c|}{\multirow{1}[0]{*}{$74.9\pm0.216$}}&\multicolumn{1}{c|}{\multirow{1}[0]{*}{$78.9\pm0.974$}}&\multicolumn{1}{c|}{\multirow{1}[0]{*}{$74.7\pm0.228$}}&&\\
\hline
\multicolumn{1}{|c|}{\multirow{6}[0]{*}{Severe}}&\multicolumn{1}{c|}{\multirow{3}[0]{*}{X-ray}}&\multicolumn{1}{c|}{\multirow{1}[0]{*}{Non-imaging info. only}}&\multicolumn{1}{c|}{\multirow{1}[0]{*}{$51.3\pm0.744$}}&\multicolumn{1}{c|}{\multirow{1}[0]{*}{$89.4\pm0.864$}}&\multicolumn{1}{c|}{\multirow{1}[0]{*}{$35.8\pm0.925$}}&\multicolumn{1}{c|}{\multirow{6}[0]{*}{141}}&\multicolumn{1}{c|}{\multirow{6}[0]{*}{57}}\\
&&\multicolumn{1}{c|}{\multirow{1}[0]{*}{Image only}}&\multicolumn{1}{c|}{\multirow{1}[0]{*}{$32.1\pm0.714$}}&\multicolumn{1}{c|}{\multirow{1}[0]{*}{$94.5\pm0.735$}}&\multicolumn{1}{c|}{\multirow{1}[0]{*}{$7.2\pm0.467$}}&&\\
&&\multicolumn{1}{c|}{\multirow{1}[0]{*}{Integrated Model}}&\multicolumn{1}{c|}{\multirow{1}[0]{*}{$60.5\pm0.775$}}&\multicolumn{1}{c|}{\multirow{1}[0]{*}{$64.0\pm1.57$}}&\multicolumn{1}{c|}{\multirow{1}[0]{*}{$59.0\pm0.959$}}&&\\
\cline{2-6}
&\multicolumn{1}{c|}{\multirow{3}[0]{*}{MRI}}&\multicolumn{1}{c|}{\multirow{1}[0]{*}{Non-imaging info. only}}&\multicolumn{1}{c|}{\multirow{1}[0]{*}{$51.3\pm0.744$}}&\multicolumn{1}{c|}{\multirow{1}[0]{*}{$89.4\pm0.864$}}&\multicolumn{1}{c|}{\multirow{1}[0]{*}{$35.8\pm0.925$}}&&\\
&&\multicolumn{1}{c|}{\multirow{1}[0]{*}{Image only}}&\multicolumn{1}{c|}{\multirow{1}[0]{*}{$34.6\pm0.775$}}&\multicolumn{1}{c|}{\multirow{1}[0]{*}{$98.3\pm0.390$}}&\multicolumn{1}{c|}{\multirow{1}[0]{*}{$9.2\pm0.632$}}&&\\
&&\multicolumn{1}{c|}{\multirow{1}[0]{*}{Integrated Model}}&\multicolumn{1}{c|}{\multirow{1}[0]{*}{$59.6\pm0.770$}}&\multicolumn{1}{c|}{\multirow{1}[0]{*}{$84.0\pm1.03$}}&\multicolumn{1}{c|}{\multirow{1}[0]{*}{$49.6\pm1.04$}}&&\\
\hline
\multicolumn{1}{|c|}{\multirow{6}[0]{*}{All}}&\multicolumn{1}{c|}{\multirow{3}[0]{*}{X-ray}}&\multicolumn{1}{c|}{\multirow{1}[0]{*}{Non-imaging info. only}}&\multicolumn{1}{c|}{\multirow{1}[0]{*}{$81.1\pm0.118$}}&\multicolumn{1}{c|}{\multirow{1}[0]{*}{$75.6\pm0.776$}}&\multicolumn{1}{c|}{\multirow{1}[0]{*}{$81.2\pm0.122$}}&\multicolumn{1}{c|}{\multirow{6}[0]{*}{5,089}}&\multicolumn{1}{c|}{\multirow{6}[0]{*}{152}}\\
&&\multicolumn{1}{c|}{\multirow{1}[0]{*}{Image only}}&\multicolumn{1}{c|}{\multirow{1}[0]{*}{$86.4\pm0.095$}}&\multicolumn{1}{c|}{\multirow{1}[0]{*}{$72.5\pm0.864$}}&\multicolumn{1}{c|}{\multirow{1}[0]{*}{$86.9\pm0.095$}}&&\\
&&\multicolumn{1}{c|}{\multirow{1}[0]{*}{Integrated Model}}&\multicolumn{1}{c|}{\multirow{1}[0]{*}{$88.4\pm0.094$}}&\multicolumn{1}{c|}{\multirow{1}[0]{*}{$66.3\pm0.924$}}&\multicolumn{1}{c|}{\multirow{1}[0]{*}{$89.1\pm0.090$}}&&\\
\cline{2-6}
&\multicolumn{1}{c|}{\multirow{3}[0]{*}{MRI}}&\multicolumn{1}{c|}{\multirow{1}[0]{*}{Non-imaging info. only}}&\multicolumn{1}{c|}{\multirow{1}[0]{*}{$81.1\pm0.118$}}&\multicolumn{1}{c|}{\multirow{1}[0]{*}{$75.6\pm0.776$}}&\multicolumn{1}{c|}{\multirow{1}[0]{*}{$81.2\pm0.122$}}&&\\
&&\multicolumn{1}{c|}{\multirow{1}[0]{*}{Image only}}&\multicolumn{1}{c|}{\multirow{1}[0]{*}{$82.1\pm0.118$}}&\multicolumn{1}{c|}{\multirow{1}[0]{*}{$84.9\pm0.636$}}&\multicolumn{1}{c|}{\multirow{1}[0]{*}{$82.1\pm0.119$}}&&\\
&&\multicolumn{1}{c|}{\multirow{1}[0]{*}{Integrated Model}}&\multicolumn{1}{c|}{\multirow{1}[0]{*}{$78.5\pm0.134$}}&\multicolumn{1}{c|}{\multirow{1}[0]{*}{$81.8\pm0.643$}}&\multicolumn{1}{c|}{\multirow{1}[0]{*}{$78.4\pm0.138$}}&&\\
\hline
\end{tabular}
\caption{\label{tab:model_perf}Performance of X-ray and MRI architectures on test data. While integrated X-ray pipeline delivers higher accuracy than integrated MRI pipeline, integrated MRI pipeline yields improved sensitivity over integrated X-ray pipeline across all stages of OA, markedly so at no OA. Standard errors used to calculate confidence intervals.}
\end{table}

\subsection*{MRI pipeline optimization and performance}
Similarly, a hyperparameter search was carried out for the MRI pipeline to optimize parameters for eventual fine-tuning. The following hyperparameters were found optimal: learning rate of \num{1.906e-2}, TKR class cross-entropy weight of 0.902 and non-TKR class weight of 0.098, dropout rate of 0.329, only last layer of model fine-tuned after OA pretrain.

An MRI-based model was fine-tuned from these parameters. The resulting predictions were fed into an LR ensemble, where averaging predictions of the best 4 models in each OA category optimized validation performance. Performance of the resulting architecture on test data is reported in the same manner as the radiograph pipeline, in Table \ref{tab:model_perf} and Fig. \ref{fig:xray_mri_performance}b. AUCs are as follows: $0.886 \pm 0.020$ (image only), $0.868 \pm 0.028$ (non-imaging info. only), $0.834 \pm 0.036$ (integrated model). AUCs for the image-only and combined pipeline versions at no OA are as follows: $0.897 \pm 0.039$ (image only); $0.943 \pm 0.029$ (integrated model). At moderate OA: $0.764 \pm 0.020$ (image only); $0.830 \pm 0.024$ (integrated model). At severe OA: $0.560 \pm 0.042$ (image only); $0.726 \pm 0.038$ (integrated model). Again, all AUC intervals are calculated using s.d., p < 0.05.

\subsection*{Comparison of MRI and radiograph pipeline performances}
\begin{figure}[h!]
\centering
\includegraphics[width=\linewidth]{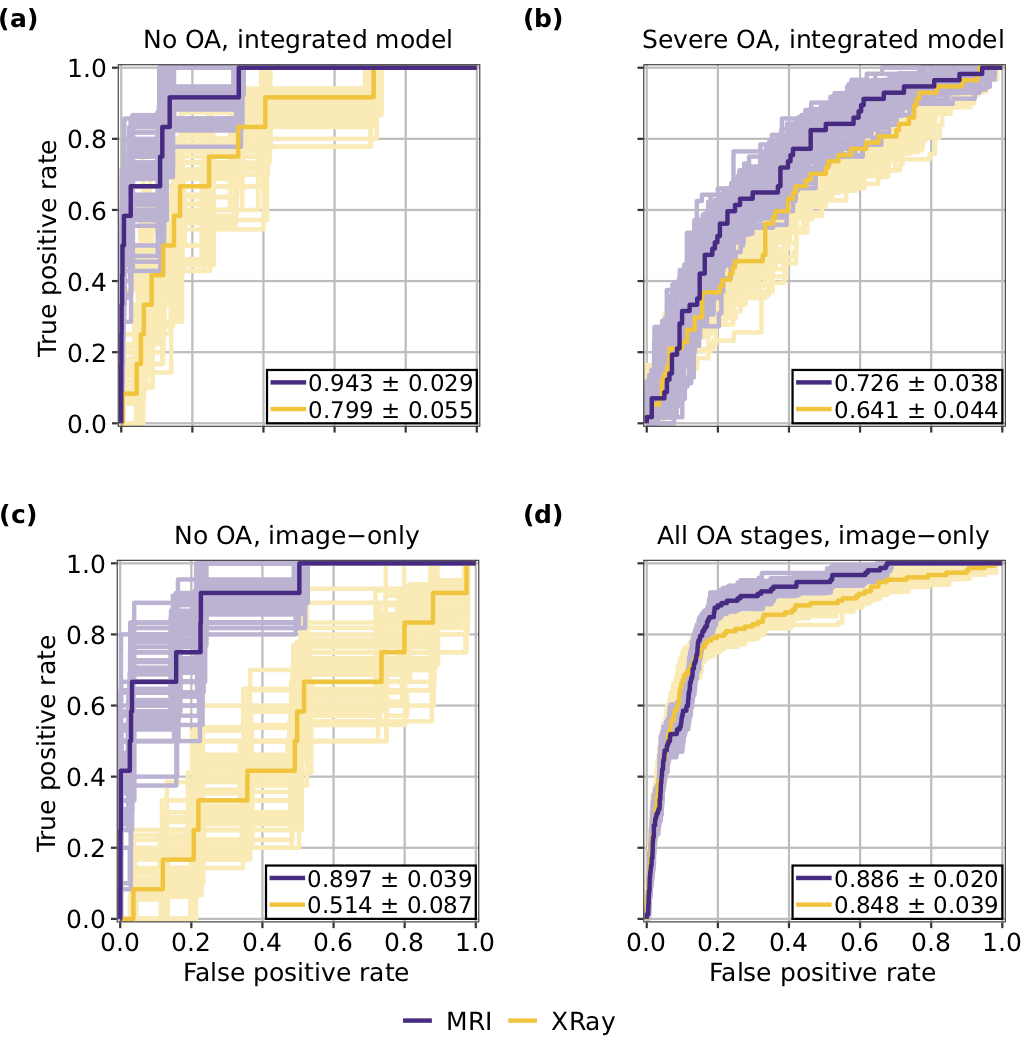}
\caption{ROC curves for MRI and X-ray pipelines at selected OA classifications and pipeline versions in which MRI performance was significantly better than that of X-ray. MRI pipeline outperforms X-ray pipeline at no OA for both image-only and integrated models, as seen in \textbf{(a)} and \textbf{(c)}. As shown in \textbf{(b)}, integrated MRI pipeline also outperformed integrated X-ray pipeline for severe OA patients, while \textbf{(d)} shows image-only MRI pipeline outperformed image-only X-ray pipeline across all OA stages. AUCs are displayed in the figure with p < 0.05. Standard deviations used to calculate confidence intervals. ROC curves with AUCs within 1 standard deviation of the mean for each pipeline version during bootstrapping are shown on plots.}
\label{fig:xray_mri_subplots}
\end{figure}

A comparison of overall AUCs attained by the integrated MRI and X-ray pipelines across OA grades and overall shows that at no OA and severe OA, the MRI pipeline outperformed the X-ray pipeline (No OA, B = 100: p = \num{3.04e-2}; Moderate OA, B = 100: p = \num{9.55e-1}; Severe OA, B = 100: p = \num{4.57e-2}; Overall, B = 100: p = \num{9.94e-1}). The MRI pipeline thus has a superior combination of sensitivity and specificity than does the X-ray pipeline for patients without OA and those with severe OA. The AUCs obtained by the image-only pipelines also were compared, and showed the MRI pipeline to outperform the X-ray pipeline for patients without OA and overall (No OA, B = 100: p = \num{6.10e-5}; Moderate OA, B = 100: p = \num{7.58e-1}; Severe OA, B = 100: p = \num{4.37e-1}; Overall, B = 100: p = \num{1.16e-2}). These results follow intuition: while radiographic imaging is primarily capable of illuminating bones in the joint, MRI can visualize soft tissues such as cartilage, muscle, and meniscus \cite{Roos2016, Menashe2012}. It follows that an MRI model will exhibit a better combination of sensitivity and specificity, especially in early OA stages at which few radiographic changes in the knee have occurred. ROC curves for pipeline versions and OA classifications in which the MRI architecture yielded a significantly better AUC than its X-ray counterpart are shown in Fig. \ref{fig:xray_mri_subplots}.

McNemar’s test assessed relative accuracies of these pipelines. There was a statistically significant difference between the accuracies of the integrated X-ray and MRI pipelines for patients at no OA, moderate OA, and overall (No OA, n = 537: p = \num{1.65e-59}; Moderate OA, n = 521: p = \num{1.13e-9}; Severe OA, n = 47: p = \num{8.84e-1}; Overall, n = 1,105: p = \num{1.52e-54}), and in each of those 3 statistically significant cases, the X-ray pipeline outperformed the MRI pipeline (No OA, n= 537: p = \num{1.11e-16}; Moderate OA, n = 521: p = \num{5.97e-10}; Overall, n = 1,105: p = \num{1.11e-16}). In interpreting these tests and the AUC tests holistically, it is evident that the X-ray pipeline is able to attain superior accuracy in several OA classifications by compromising on its combination of sensitivity and specificity. This is further supported by the accuracies and sensitivities reported for the respective pipelines in Table \ref{tab:model_perf}, which show that while the X-ray pipeline is more accurate than its MRI counterpart at every OA classification, the opposite is true for sensitivity—drastically so for patients without OA. In the clinic, where sensitivity as to whether a patient is at risk of eventual TKR is paramount, these results would show the MRI pipeline to be the more useful model.

It is also worthy to note the improvement in performance that occurs for patients without OA when imaging predictions are added to non-imaging variables in both pipelines. In the X-ray pipeline, the model’s AUC increased from $0.514 \pm 0.087$ to $0.799 \pm 0.055$ when non-imaging variables were added to the radiographs, a sizeable increase when compared to the MRI pipeline performance, which saw AUC increase from $0.897 \pm 0.039$ to $0.943 \pm 0.029$ (p < 0.05 for all). This demonstrates that non-imaging variables such as various pain scales seem to add critical information to the X-ray pipeline, while the same information is less important in the MRI pipeline.

\subsection*{Biomarker identification and analysis}
\begin{figure}[h]
\centering
\includegraphics[width=\linewidth]{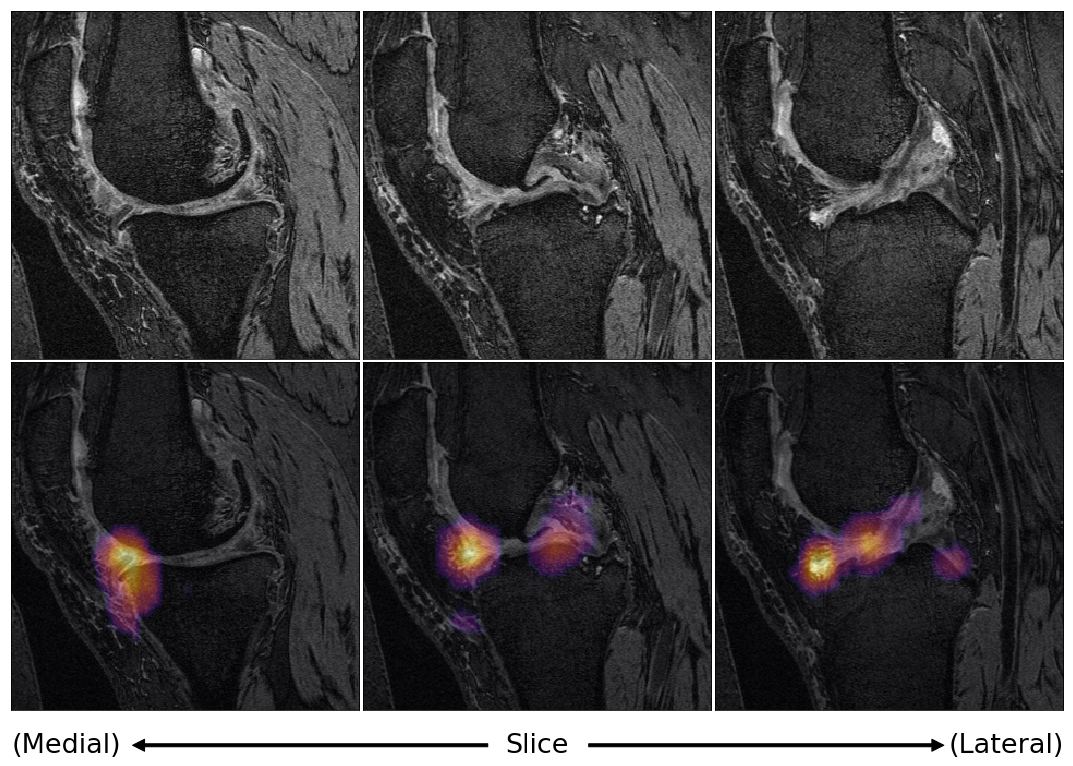}
\caption{Slices of occlusion map of true positive detected by MRI pipeline, overlaid on corresponding slices of DESS MRI. Such maps were generated and analyzed for all 124 true positives and corresponding true negative controls of the integrated MRI pipeline.}
\label{fig:sample_occ_map}
\end{figure}

Of the 152 patients in test data who underwent a TKR, 124 were detected by the MRI pipeline. Occlusion maps were generated for these cases and their corresponding true negative controls, an example of which is shown in Fig. \ref{fig:sample_occ_map}. Tissues and their hotspot percentages across these true positives and corresponding true negative controls can be found in Supplementary Table \ref{tab:occ_map_tp_summary} and Supplementary Table \ref{tab:occ_map_tn_summary}, respectively. ORs, 95\% confidence intervals, and associated p values for each tissue can be found in Table \ref{tab:occ_map_analysis}.

\begin{table}[ht]
\centering
\begin{tabular}{ |P{0.7cm}|P{0.3cm}|P{0.8cm}|P{1.9cm}|} 
\hline
\multicolumn{1}{|c|}{\multirow{1}{*}{\textbf{Tissue type}}}&\multicolumn{1}{c|}{\multirow{1}{*}{\textbf{Tissue}}}&\multicolumn{1}{c|}{\multirow{1}{*}{\textbf{OR (95\% CI, n = 124)}}}&\multicolumn{1}{c|}{\multirow{1}{*}{\textbf{p value (n = 124)}}}\\
\hline
\multicolumn{1}{|c|}{\multirow{3}{*}{Cartilage}}&\multicolumn{1}{c|}{TFJ medial*}&\multicolumn{1}{c|}{0.05 (0.00 - 0.48)} & \multicolumn{1}{c|}{\num{3.36e-3}}\\
&\multicolumn{1}{c|}{TFJ lateral*}&\multicolumn{1}{c|}{0.03 (0.00 - 0.25)} & \multicolumn{1}{c|}{\num{3.89e-5}}\\
&\multicolumn{1}{c|}{PFJ}&\multicolumn{1}{c|}{1.03 (0.60 - 1.77)} & \multicolumn{1}{c|}{\num{1.00e0}}\\
\hline
\multicolumn{1}{|c|}{\multirow{4}{*}{Meniscus}}&\multicolumn{1}{c|}{Medial anterior*}&\multicolumn{1}{c|}{0.33 (0.12 - 0.79)} & \multicolumn{1}{c|}{\num{1.04e-2}}\\
&\multicolumn{1}{c|}{Medial posterior*}&\multicolumn{1}{c|}{0.40 (0.16 - 0.89)} & \multicolumn{1}{c|}{\num{2.37e-2}}\\
&\multicolumn{1}{c|}{Lateral anterior*}&\multicolumn{1}{c|}{0.23 (0.05 - 0.67)} & \multicolumn{1}{c|}{\num{5.05e-3}}\\&\multicolumn{1}{c|}{Lateral posterior*}&\multicolumn{1}{c|}{0.26 (0.06 - 0.80)} & \multicolumn{1}{c|}{\num{1.49e-2}}\\
\hline
\multicolumn{1}{|c|}{\multirow{3}{*}{Bone}}&\multicolumn{1}{c|}{TFJ medial*}&\multicolumn{1}{c|}{0.17 (0.00 - 0.91)} & \multicolumn{1}{c|}{\num{3.57e-2}}\\
&\multicolumn{1}{c|}{TFJ lateral*}&\multicolumn{1}{c|}{0.02 (0.00 - 0.22)} & \multicolumn{1}{c|}{\num{8.48e-6}}\\
&\multicolumn{1}{c|}{PFJ}&\multicolumn{1}{c|}{1.11 (0.64 - 1.92)} & \multicolumn{1}{c|}{\num{7.93e-1}}\\
\hline
\multicolumn{1}{|c|}{\multirow{3}{*}{Ligament}}&\multicolumn{1}{c|}{ACL*}&\multicolumn{1}{c|}{0.49 (0.23 - 0.99)} & \multicolumn{1}{c|}{\num{4.72e-2}}\\
&\multicolumn{1}{c|}{PCL}&\multicolumn{1}{c|}{1.58 (0.89 - 2.87)} & \multicolumn{1}{c|}{\num{1.27e-1}}\\
&\multicolumn{1}{c|}{Popliteal}&\multicolumn{1}{c|}{1.62 (0.96 - 2.77)} & \multicolumn{1}{c|}{\num{7.51e-2}}\\
\hline
\multicolumn{1}{|c|}{\multirow{8}{*}{Tendon}}&\multicolumn{1}{c|}{Medial patellar retinaculum*}&\multicolumn{1}{c|}{1.98 (1.02 - 3.99)} & \multicolumn{1}{c|}{\num{4.19e-2}}\\
&\multicolumn{1}{c|}{Lateral patellar retinaculum}&\multicolumn{1}{c|}{1.08 (0.60 - 1.96)} & \multicolumn{1}{c|}{\num{8.88e-1}}\\
&\multicolumn{1}{c|}{Popliteal}&\multicolumn{1}{c|}{1.49 (0.87 - 2.57)} & \multicolumn{1}{c|}{\num{1.56e-1}}\\&\multicolumn{1}{c|}{Patellar}&\multicolumn{1}{c|}{1.76 (0.92 - 3.48)} & \multicolumn{1}{c|}{\num{9.00e-2}}\\&\multicolumn{1}{c|}{Gastrocnemius*}&\multicolumn{1}{c|}{2.97 (1.12 - 10.0)} & \multicolumn{1}{c|}{\num{2.67e-2}}\\&\multicolumn{1}{c|}{Semimembranosus}&\multicolumn{1}{c|}{0.50 (0.23 - 1.03)} & \multicolumn{1}{c|}{\num{6.17e-2}}\\&\multicolumn{1}{c|}{Quadriceps}&\multicolumn{1}{c|}{3.18 (0.88 - 20.4)} & \multicolumn{1}{c|}{\num{8.38e-2}}\\&\multicolumn{1}{c|}{Gracilis}&\multicolumn{1}{c|}{4.52 (0.74 - 290)} & \multicolumn{1}{c|}{\num{1.20e-1}}\\
\hline
\multicolumn{1}{|c|}{\multirow{1}{*}{Fat Pad}}&\multicolumn{1}{c|}{Hoffa}&\multicolumn{1}{c|}{2.38 (0.92 - 7.38)} & \multicolumn{1}{c|}{\num{7.80e-2}}\\
\hline
\multicolumn{1}{|c|}{\multirow{7}{*}{Muscle}}&\multicolumn{1}{c|}{Popliteus}&\multicolumn{1}{c|}{1.98 (1.00 - 4.14)} & \multicolumn{1}{c|}{\num{5.11e-2}}\\
&\multicolumn{1}{c|}{Vastus medialis}&\multicolumn{1}{c|}{1.26 (0.54 - 3.00)} & \multicolumn{1}{c|}{\num{6.93e-1}}\\
&\multicolumn{1}{c|}{Gastrocnemius}&\multicolumn{1}{c|}{1.35 (0.73 - 2.54)} & \multicolumn{1}{c|}{\num{3.76e-1}}\\&\multicolumn{1}{c|}{Plantaris*}&\multicolumn{1}{c|}{2.84 (1.47 - 5.82)} & \multicolumn{1}{c|}{\num{1.29e-3}}\\&\multicolumn{1}{c|}{Biceps femoris}&\multicolumn{1}{c|}{4.52 (0.74 - 290)} & \multicolumn{1}{c|}{\num{1.20e-1}}\\&\multicolumn{1}{c|}{Tibilais anterior}&\multicolumn{1}{c|}{2.37 (0.24 - 161)} & \multicolumn{1}{c|}{\num{6.22e-1}}\\&\multicolumn{1}{c|}{Semimembranosus}&\multicolumn{1}{c|}{0.35 (0.05 - 1.32)} & \multicolumn{1}{c|}{\num{1.36e-1}}\\
\hline
\multicolumn{1}{|c|}{\multirow{1}{*}{Synovium}}&\multicolumn{1}{c|}{General}&\multicolumn{1}{c|}{1.17 (0.50 - 2.82)} & \multicolumn{1}{c|}{\num{8.41e-1}}\\
\hline
\end{tabular}
\caption{\label{tab:occ_map_analysis}Summary of occlusion map analysis comparing hotspot frequencies in selected knee joint tissues. Hotspots were defined as pixels that, when occluded, were in the top 5\% of all pixels in change of pipeline TKR prediction output metric. Odds ratios, 95\% confidence intervals calculated using Cornfield’s method, and p values calculated using Fisher’s exact test are displayed. Tissues significant at $\alpha = 0.05$ are designated with a *. n = 124 for all tests.}
\end{table}

Three tissues saw ORs and 95\% confidence intervals that lied above 1 and p values below $\alpha = 0.05$: the medial patellar retinaculum, gastrocnemius tendon, and plantaris muscle. Thus, we conclude there is a substantial and statistically significant difference in the risk of TKR within 5 years when these tissues are identified as hotspots by the pipeline. From the ORs, we see that the risk of TKR increases when any of the three are identified as hotspots: for the medial patellar retinaculum, the risk is 1.98 times higher with a 95\% confidence interval from 1.02 to 3.99; for the gastrocnemius tendon, it is 2.97 times higher with a 95\% confidence interval from 1.12 to 10.0; and for the plantaris muscle, it is 2.84 times higher with a 95\% confidence interval from 1.47 to 5.82. As such, these results provide evidence that all are imaging biomarkers of TKR.

On the other hand, several tissues located within or near the tibiofemoral joint—namely, cartilage and bone in both medial and lateral locations of the joint, menisci in all tested regions, and the ACL—saw ORs and 95\% confidence intervals entirely below 1 and p values below $\alpha = 0.05$. Consequently, for all of these tissues, we find a statistically significant difference in the risk of TKR within 5 years when these tissues are identified as hotspots. In the case of each, the risk of TKR appears to decrease when these tissues are identified as hotspots. Interestingly, each of these tissues have either been implicated as imaging biomarkers of OA progression, or damage within them is associated with OA onset \cite{Collins2016, Khan2019, Simon2015}. These results, in conjunction with the three tissues in which risk of TKR increased when identified as hotspots, suggest that compared to OA progression, TKR onset relies less on tissues in and around the tibiofemoral joint and more on tissues in other locations of the joint to make predictions. TKR has been considered an outcome of OA progression, but these results demonstrate in part how it is a more nuanced problem.

\section*{Discussion}
In this work, we present a pipeline that integrates MR imaging and non-imaging features to attain strong TKR prediction performance, reporting accuracy of $78.5 \pm 0.134\%$, sensitivity of $81.8 \pm 0.643\%$, and specificity of $78.4 \pm 0.138\%$ (intervals calculated with standard error of  measurement (s.e.m.), p < 0.05). Comparisons of AUCs showed the MRI pipeline to outperform the X-ray pipeline for patients without OA and with severe OA, thereby showing the MRI model to have a better combination of sensitivity and specificity in these OA classifications. That it did so particularly for patients without OA shows the utility of the MRI pipeline in screening for patients at risk of TKR despite higher costs. It was also interesting that, particularly among patients with no OA, the X-ray model improved drastically more than the MRI model when non-imaging information was added, judging by disparities in AUCs. This suggests the MRI-trained DenseNet-121 may have learned to predict some of the non-imaging features from the images themselves, indicating that MRI images may intrinsically contain information regarding pain, quality of life, and physical performance, among other non-imaging variables used in this study. The utility of MRI in predicting these variables through DL is certainly worth further investigation.

A comparison of the MRI pipeline performance to past work is insightful. The closest analog to our work was conducted by Wang, T. \emph{et al.} \cite{Wang2019}, who trained independent residual networks to predict TKR from both DESS and Turbo Spin Echo (TSE) MRI images, integrating both predictions with non-imaging variables in a LR model to yield a final TKR prediction. This yielded a model with AUC of $0.86 \pm 0.01$ (p < 0.01) when solely DESS or TSE images were used, and $0.88 \pm 0.02$ (p < 0.01) when both images and non-imaging features were integrated. Our MRI image-only model saw AUC of $0.886 \pm 0.020$ (image only, p < 0.05) and an integrated AUC of $0.834 \pm 0.036$ (combined, p < 0.05). Our image-only model thus yields performance superior to its image-only counterpart, with a 95\% confidence interval lying entirely above the mean AUC of the image-only model by Wang, T. \emph{et al.} \cite{Wang2019}. Our integrated model, as discussed previously, was optimized to maximize Youden’s index within each OA classification rather than overall AUC, explaining why our integrated model has a lower overall AUC than our image-only model. However, due to this decision, we obtained strong performance at early and moderate OA stages, with sensitivity and specificity of $92.2 \pm 1.68\%$ and $82.4 \pm 0.173\%$ at no OA, respectively, and $78.9 \pm 0.974\%$ and $74.7 \pm 0.228\%$ at moderate OA (intervals calculated using s.e.m., p < 0.05). In particular, the AUC of $0.943 \pm 0.029$ (interval calculated with s.d., p < 0.05) obtained by the MRI pipeline for patients without OA, the most difficult OA classification from which to predict TKR, by far surpasses that of past TKR predictive models that include patients across all stages of OA. This performance marks progress towards a model that identifies patients at risk for TKR such that nonsurgical treatment strategies can be implemented to delay TKR.

The biomarker analysis conducted also has implications, as it identified several tissues located within or near the tibiofemoral joint as reducing risk of TKR when identified as hotspots by the full MRI pipeline—namely, these were medially and laterally located cartilage and bone, all examined meniscal regions, and the ACL. These tissues or damage within them all have been associated with progression or onset of OA, and that our model shows TKR onset to be less reliant on these imaging features in cases compared to controls demonstrates TKR onset to be a more complicated problem than OA progression, despite the relationship between the two. On the other hand, the model identifies three tissues as increasing risk of TKR when identified as hotspots in the pipeline: the medial patellar retinaculum, gastrocnemius tendon, and plantaris muscle. The medial patellar retinaculum is crucial for lateral stabilization of the knee joint, and as such, damage to it results in a patella that more easily dislocates \cite{Diederichs2010}. Past work has shown patellar dislocation increases risk for OA, and TKR can be an effective procedure to treat inveterate patellar dislocation, showing a previous link between this tissue’s functionality and eventual OA and TKR \cite{Moatshe2017, Figueroa2019}. The gastrocnemius tendon and plantaris muscle, on the other hand, are both posteriorly located tissues within the knee that play a key role in knee flexion \cite{Souza2010}. While literature regarding the plantaris muscle is rather sparse, injuries to the muscle can be implicated in knee and calf pain felt by a patient \cite{Spina2007}. Given their related functionality and location, the gastrocnemius tendon and plantaris muscle can jointly be implicated in conditions such as “tennis leg,” which refers to mid-calf pain felt during extension of the leg, usually due to damage to one of these tissues or their associated muscles or tendons \cite{Zetaruk2007}. The significance of the plantaris muscle and gastrocnemius tendon to OA progression and TKR, however, have not been well characterized, and these results justify future studies to these ends.

This study had some limitations. The first is specific to the OAI dataset, which tends towards older, female patients, all from the United States: across 4,796 patients, the mean age is 61 years and 58\% of patients are female. This is not emblematic of the general population, so the robustness of the pipeline could be strengthened by testing on a dataset such as the Multicenter Osteoarthritis Study (MOST). A further limitation of the dataset is that, despite the fairly large size, there are a very limited number of patients with the classification of most interest: those without radiographic OA that still undergo TKR within 5 years. Only 66 such cases existed in the entire OAI dataset, and 12 were in the test set. As such, the OAI dataset and the number of comparison experiments we ran within and across OA classifications limits the statistical power of our conclusions. Furthermore, in this study, pixels in MRI images were compressed to 14 possible values to optimize performance—a version of the pipeline was also constructed and evaluated without the compression, but its TKR prediction performance was not as strong. Ideally, a model that uses all available information would be used in occlusion map analysis to draw more precise conclusions regarding anatomic regions that associate with TKR, but this compromise was necessary to improve performance. A final limitation was computational intensity in occlusion map generation: the voxel size and stride used were $12 \times 32 \times 32$ and 12, respectively. These ideally would be smaller so maps could yield more precise insights but doing so was infeasible in a reasonable amount of time.

To conclude, this work presents a predictive model that delivers performance not previously seen in predicting TKR, especially for patients without OA. By delivering such performance, this pipeline can identify patients at risk of TKR with high sensitivity and specificity, and for patients with no or moderate OA, this can allow a non-invasive treatment to be implemented that prolongs good health of the knee and delays TKR. The biomarker analysis identifies the medial patellar retinaculum, gastrocnemius tendon, and plantaris muscle as increasing risk of TKR when identified as a hotspot by the model, while its assessment that several tissues within and near the tibiofemoral joint appear to reduce risk of TKR helps demonstrate the added complexity of predicting TKR onset as opposed to OA progression. Beyond this, additional directions include investigating a more effective means of integrating non-image information with image predictions to improve TKR prediction performance, assessing the efficacy of alternate network architectures, and reducing computational time to make predictions. 

\section*{Acknowledgements}

The OAI is a public-private partnership comprised of five contracts (N01-AR-2-2258; N01-AR-2-2259; N01-AR-2-2260; N01-AR-2-2261; N01-AR-2-2262) funded by the National Institutes of Health, a branch of the Department of Health and Human Services, and conducted by the OAI Study Investigators. Private funding partners include Merck Research Laboratories; Novartis Pharmaceuticals Corporation, GlaxoSmithKline; and Pfizer, Inc. Private sector funding for the OAI is managed by the Foundation for the National Institutes of Health. This manuscript was prepared using an OAI public use data set and does not necessarily reflect the opinions or views of the OAI investigators, the NIH, or the private funding partners. This study was supported by the grants R61AR073552 (S.M./V.P.) and R00AR070902 (V.P.). Institutional research funds are provided by GE Healthcare for unrelated studies.

\section*{Additional information}

\subsection*{Competing interests}
The authors declare no competing interests.

\bibliography{sample}
\section*{Supplementary Figures}

\begin{figure}[h]
\centering
\includegraphics[width={6.7 in}]{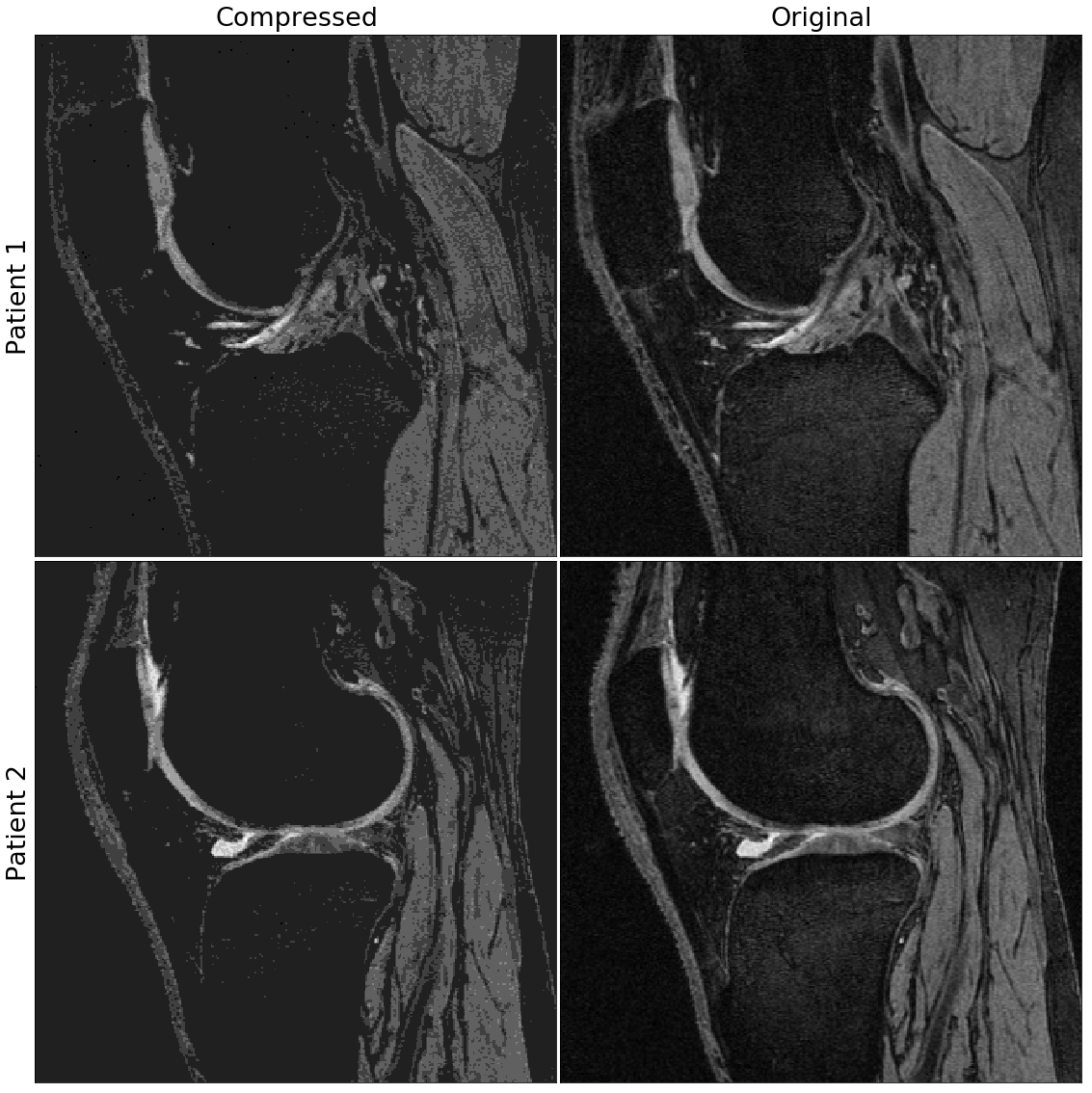}
\caption{Sample slices of DESS MRI and their corresponding compressed versions when rounding pixel values after normalization.}
\label{fig:image_compression}
\end{figure}

\section*{Supplementary Tables}

\begin{table}[h]
\centering
\begin{tabular}{ |P{0.7cm}|P{0.3cm}|P{0.8cm}|} 
\hline
\multicolumn{1}{|c|}{\multirow{1}{*}{\textbf{Variable grouping}}}&\multicolumn{1}{c|}{\multirow{1}{*}{\textbf{Variable}}}&\multicolumn{1}{c|}{\multirow{1}{*}{\textbf{Source}}}\\
\hline
\multicolumn{1}{|c|}{\multirow{6}{*}{Demographics}}&\multicolumn{1}{l|}{Age}&\multicolumn{1}{c|}{(Lewis, 2013) \cite{Lewis2013}}\\
&\multicolumn{1}{l|}{Obesity/BMI}&\multicolumn{1}{c|}{(Lewis, 2013) \cite{Lewis2013}}\\
&\multicolumn{1}{l|}{Gender}&\multicolumn{1}{c|}{(Heidari, 2011) \cite{Heidari2011}}\\
&\multicolumn{1}{l|}{Ethnicity}&\multicolumn{1}{c|}{(Yu, 2019) \cite{Yu2019}}\\
&\multicolumn{1}{l|}{Income}&\multicolumn{1}{c|}{(Hawker, 2006) \cite{Hawker2006}}\\
&\multicolumn{1}{l|}{Education level}&\multicolumn{1}{c|}{(Pisters, 2012) \cite{Pisters2012}}\\
\hline
\multicolumn{1}{|c|}{\multirow{5}{*}{Previous knee trauma and pain}}&\multicolumn{1}{l|}{Knee pain}&\multicolumn{1}{c|}{(Lewis, 2013) \cite{Lewis2013}}\\
&\multicolumn{1}{l|}{Previous knee trauma}&\multicolumn{1}{c|}{(Heidari, 2011) \cite{Heidari2011}}\\
&\multicolumn{1}{l|}{Repetitive knee trauma}&\multicolumn{1}{c|}{(Heidari, 2011) \cite{Heidari2011}}\\
&\multicolumn{1}{l|}{Previous meniscal injuries}&\multicolumn{1}{c|}{(Heidari, 2011) \cite{Heidari2011}}\\
&\multicolumn{1}{l|}{Previous knee injury}&\multicolumn{1}{c|}{(Cooper, 2000) \cite{Cooper2000}}\\
\hline
\multicolumn{1}{|c|}{\multirow{7}{*}{Knee physical activity and functionality}}&\multicolumn{1}{l|}{Mechanical forces exerted on knee}&\multicolumn{1}{c|}{(Heidari, 2011) \cite{Heidari2011}}\\
&\multicolumn{1}{l|}{Frequent kneeling}&\multicolumn{1}{c|}{(Heidari, 2011) \cite{Heidari2011}}\\
&\multicolumn{1}{l|}{Frequent squatting}&\multicolumn{1}{c|}{(Heidari, 2011) \cite{Heidari2011}}\\
&\multicolumn{1}{l|}{Physical activity level}&\multicolumn{1}{c|}{(Pisters, 2012) \cite{Pisters2012}}\\
&\multicolumn{1}{l|}{Muscular weakness}&\multicolumn{1}{c|}{(Heidari, 2011) \cite{Heidari2011}}\\
&\multicolumn{1}{l|}{Joint range of motion}&\multicolumn{1}{c|}{(Pisters, 2012) \cite{Pisters2012}}\\
&\multicolumn{1}{l|}{Lower knee extension muscle strength}&\multicolumn{1}{c|}{(Pisters, 2012) \cite{Pisters2012}}\\
\hline
\multicolumn{1}{|c|}{\multirow{8}{*}{Previous actions to treat knee pain}}&\multicolumn{1}{l|}{Previous joint injections}&\multicolumn{1}{c|}{(Yu, 2019) \cite{Yu2019}}\\
&\multicolumn{1}{l|}{Previous knee arthroscopy}&\multicolumn{1}{c|}{(Yu, 2019) \cite{Yu2019}}\\
&\multicolumn{1}{l|}{Previous analgesics or opioid usage}&\multicolumn{1}{c|}{(Lewis, 2013) \cite{Lewis2013}}\\
&\multicolumn{1}{l|}{Previous NSAID usage}&\multicolumn{1}{c|}{(Yu, 2019) \cite{Yu2019}}\\
&\multicolumn{1}{l|}{Number of previous knee referrals}&\multicolumn{1}{c|}{(Yu, 2019) \cite{Yu2019}}\\
&\multicolumn{1}{l|}{Number of previous consultations}&\multicolumn{1}{c|}{(Yu, 2019) \cite{Yu2019}}\\
&\multicolumn{1}{l|}{Willingness to consider TJA as treatment}&\multicolumn{1}{c|}{(Hawker, 2006) \cite{Hawker2006}}\\
&\multicolumn{1}{l|}{Seen physician for arthritis in previous year}&\multicolumn{1}{c|}{(Hawker, 2006) \cite{Hawker2006}}\\
\hline
\multicolumn{1}{|c|}{\multirow{10}{*}{Preexisting health conditions}}&\multicolumn{1}{l|}{Heberden's nodes}&\multicolumn{1}{c|}{(Cooper, 2000) \cite{Cooper2000}}\\
&\multicolumn{1}{l|}{Recorded diagnosis of joint-specific OA}&\multicolumn{1}{c|}{(Yu, 2019) \cite{Yu2019}}\\
&\multicolumn{1}{l|}{Low back pain}&\multicolumn{1}{c|}{(Yu, 2019) \cite{Yu2019}}\\
&\multicolumn{1}{l|}{Hypertension}&\multicolumn{1}{c|}{(Yu, 2019) \cite{Yu2019}}\\
&\multicolumn{1}{l|}{Smoking status}&\multicolumn{1}{c|}{(Yu, 2019) \cite{Yu2019}}\\
&\multicolumn{1}{l|}{Drinking status}&\multicolumn{1}{c|}{(Yu, 2019) \cite{Yu2019}}\\
&\multicolumn{1}{l|}{Asthma}&\multicolumn{1}{c|}{(Yu, 2019) \cite{Yu2019}}\\
&\multicolumn{1}{l|}{COPD}&\multicolumn{1}{c|}{(Yu, 2019) \cite{Yu2019}}\\
&\multicolumn{1}{l|}{Diabetes mellitus}&\multicolumn{1}{c|}{(Yu, 2019) \cite{Yu2019}}\\
&\multicolumn{1}{l|}{Comorbidities}&\multicolumn{1}{c|}{(Pisters, 2012) \cite{Pisters2012}}\\
\hline
\multicolumn{1}{|c|}{\multirow{4}{*}{Miscellaneous}}&\multicolumn{1}{l|}{Knee joint laxity}&\multicolumn{1}{c|}{(Heidari, 2011) \cite{Heidari2011}}\\
&\multicolumn{1}{l|}{Genetic susceptibility to knee OA}&\multicolumn{1}{c|}{(Heidari, 2011) \cite{Heidari2011}}\\
&\multicolumn{1}{l|}{Mental health measures}&\multicolumn{1}{c|}{(Sharma, 2003) \cite{Sharma2003}}\\
&\multicolumn{1}{l|}{SF36 score}&\multicolumn{1}{c|}{(Hawker, 2006) \cite{Hawker2006}}\\
\hline
\end{tabular}
\caption{\label{tab:potential_variable_list}Non-imaging variables identified from literature as correlated with OA progression or eventual TKR. These non-imaging variables were taken to the OAI database, and, if present, added as potential non-imaging variables to supplement image-based predictions.}
\end{table}

\begin{table}[ht]
\centering
\begin{tabular}{ |P{0.7cm}|P{0.3cm}|P{0.8cm}|P{1.9cm}|P{1.9cm}|P{1.9cm}|} 
\hline
\multicolumn{1}{|c|}{\multirow{2}{*}{\textbf{Tissue type}}}&\multicolumn{1}{c|}{\multirow{2}{*}{\textbf{Tissue}}}&\multicolumn{1}{c|}{\multirow{2}{*}{\begin{tabular}{@{}c@{}}\textbf{No OA} \\ \textbf{(n = 11)}\end{tabular}}}&\multicolumn{1}{c|}{\multirow{2}{*}{\begin{tabular}{@{}c@{}}\textbf{Moderate OA} \\ \textbf{(n = 65)}\end{tabular}}}&\multicolumn{1}{c|}{\multirow{2}{*}{\begin{tabular}{@{}c@{}}\textbf{Severe OA} \\ \textbf{(n = 48)}\end{tabular}}}&\multicolumn{1}{c|}{\multirow{2}{*}{\begin{tabular}{@{}c@{}}\textbf{Total} \\ \textbf{(n = 124)}\end{tabular}}}\\
&&&&&\\
\hline
\multicolumn{1}{|c|}{\multirow{3}{*}{Cartilage}}&\multicolumn{1}{c|}{TFJ medial}&\multicolumn{1}{c|}{100} & \multicolumn{1}{c|}{95.4} &  \multicolumn{1}{c|}{87.5} & \multicolumn{1}{c|}{92.7}\\
&\multicolumn{1}{c|}{TFJ lateral}&\multicolumn{1}{c|}{100} & \multicolumn{1}{c|}{87.7}&\multicolumn{1}{c|}{85.4}&\multicolumn{1}{c|}{87.9}\\
&\multicolumn{1}{c|}{PFJ}&\multicolumn{1}{c|}{27.3} & \multicolumn{1}{c|}{43.1}&\multicolumn{1}{c|}{41.7}&\multicolumn{1}{c|}{41.1}\\
\hline
\multicolumn{1}{|c|}{\multirow{4}{*}{Meniscus}}&\multicolumn{1}{c|}{Medial anterior}&\multicolumn{1}{c|}{100} & \multicolumn{1}{c|}{84.6} &  \multicolumn{1}{c|}{75.0} & \multicolumn{1}{c|}{82.3}\\
&\multicolumn{1}{c|}{Medial posterior}&\multicolumn{1}{c|}{90.9} & \multicolumn{1}{c|}{87.7}&\multicolumn{1}{c|}{70.8}&\multicolumn{1}{c|}{81.5}\\
&\multicolumn{1}{c|}{Lateral anterior}&\multicolumn{1}{c|}{100} & \multicolumn{1}{c|}{87.7}&\multicolumn{1}{c|}{81.3}&\multicolumn{1}{c|}{86.3}\\
&\multicolumn{1}{c|}{Lateral posterior}&\multicolumn{1}{c|}{100} & \multicolumn{1}{c|}{90.8}&\multicolumn{1}{c|}{81.3}&\multicolumn{1}{c|}{87.9}\\
\hline
\multicolumn{1}{|c|}{\multirow{3}{*}{Bone}}&\multicolumn{1}{c|}{TFJ medial}&\multicolumn{1}{c|}{100} & \multicolumn{1}{c|}{95.4} &  \multicolumn{1}{c|}{89.6} & \multicolumn{1}{c|}{93.5}\\
&\multicolumn{1}{c|}{TFJ lateral}&\multicolumn{1}{c|}{90.9} & \multicolumn{1}{c|}{87.7}&\multicolumn{1}{c|}{83.3}&\multicolumn{1}{c|}{86.3}\\
&\multicolumn{1}{c|}{PFJ}&\multicolumn{1}{c|}{27.3} & \multicolumn{1}{c|}{35.4}&\multicolumn{1}{c|}{45.8}&\multicolumn{1}{c|}{38.7}\\
\hline
\multicolumn{1}{|c|}{\multirow{3}{*}{Ligament}}&\multicolumn{1}{c|}{ACL}&\multicolumn{1}{c|}{100} & \multicolumn{1}{c|}{81.5} &  \multicolumn{1}{c|}{64.6} & \multicolumn{1}{c|}{76.6}\\
&\multicolumn{1}{c|}{PCL}&\multicolumn{1}{c|}{72.7} & \multicolumn{1}{c|}{73.8}&\multicolumn{1}{c|}{77.1}&\multicolumn{1}{c|}{75.0}\\
&\multicolumn{1}{c|}{Popliteal}&\multicolumn{1}{c|}{54.5} & \multicolumn{1}{c|}{56.9}&\multicolumn{1}{c|}{58.3}&\multicolumn{1}{c|}{57.3}\\
\hline
\multicolumn{1}{|c|}{\multirow{8}{*}{Tendon}}&\multicolumn{1}{c|}{Medial patellar retinaculum}&\multicolumn{1}{c|}{90.9} & \multicolumn{1}{c|}{78.5} &  \multicolumn{1}{c|}{91.7} & \multicolumn{1}{c|}{84.7}\\
&\multicolumn{1}{c|}{Lateral patellar retinaculum}&\multicolumn{1}{c|}{54.5} & \multicolumn{1}{c|}{21.5}&\multicolumn{1}{c|}{33.3}&\multicolumn{1}{c|}{29.0}\\
&\multicolumn{1}{c|}{Popliteal}&\multicolumn{1}{c|}{36.4} & \multicolumn{1}{c|}{49.2}&\multicolumn{1}{c|}{43.8}&\multicolumn{1}{c|}{46.0}\\
&\multicolumn{1}{c|}{Patellar}&\multicolumn{1}{c|}{27.3} & \multicolumn{1}{c|}{27.7}&\multicolumn{1}{c|}{25.0}&\multicolumn{1}{c|}{26.6}\\
&\multicolumn{1}{c|}{Gastrocnemius}&\multicolumn{1}{c|}{36.4} & \multicolumn{1}{c|}{9.2}&\multicolumn{1}{c|}{14.6}&\multicolumn{1}{c|}{13.7}\\
&\multicolumn{1}{c|}{Semimembranosus}&\multicolumn{1}{c|}{27.3} & \multicolumn{1}{c|}{13.8}&\multicolumn{1}{c|}{6.3}&\multicolumn{1}{c|}{12.1}\\
&\multicolumn{1}{c|}{Quadriceps}&\multicolumn{1}{c|}{0.0} & \multicolumn{1}{c|}{4.6}&\multicolumn{1}{c|}{14.6}&\multicolumn{1}{c|}{8.1}\\
&\multicolumn{1}{c|}{Gracilis}&\multicolumn{1}{c|}{0.0} & \multicolumn{1}{c|}{4.6}&\multicolumn{1}{c|}{6.3}&\multicolumn{1}{c|}{4.8}\\
\hline
\multicolumn{1}{|c|}{\multirow{1}{*}{Fat Pad}}&\multicolumn{1}{c|}{Hoffa}&\multicolumn{1}{c|}{100} & \multicolumn{1}{c|}{90.8} &  \multicolumn{1}{c|}{97.9} & \multicolumn{1}{c|}{94.4}\\
\hline
\multicolumn{1}{|c|}{\multirow{7}{*}{Muscle}}&\multicolumn{1}{c|}{Popliteus}&\multicolumn{1}{c|}{18.2} & \multicolumn{1}{c|}{35.4} &  \multicolumn{1}{c|}{10.4} & \multicolumn{1}{c|}{24.2}\\
&\multicolumn{1}{c|}{Vastus medialis}&\multicolumn{1}{c|}{18.2} & \multicolumn{1}{c|}{7.7}&\multicolumn{1}{c|}{18.8}&\multicolumn{1}{c|}{12.9}\\
&\multicolumn{1}{c|}{Gastrocnemius}&\multicolumn{1}{c|}{36.4} & \multicolumn{1}{c|}{26.2}&\multicolumn{1}{c|}{27.1}&\multicolumn{1}{c|}{27.4}\\
&\multicolumn{1}{c|}{Plantaris}&\multicolumn{1}{c|}{27.3} & \multicolumn{1}{c|}{32.3}&\multicolumn{1}{c|}{31.3}&\multicolumn{1}{c|}{31.5}\\
&\multicolumn{1}{c|}{Biceps femoris}&\multicolumn{1}{c|}{0.0} & \multicolumn{1}{c|}{4.6}&\multicolumn{1}{c|}{6.3}&\multicolumn{1}{c|}{4.8}\\
&\multicolumn{1}{c|}{Tibialis anterior}&\multicolumn{1}{c|}{0.0} & \multicolumn{1}{c|}{4.6}&\multicolumn{1}{c|}{0.0}&\multicolumn{1}{c|}{2.4}\\
&\multicolumn{1}{c|}{Semimembranosus}&\multicolumn{1}{c|}{0.0} & \multicolumn{1}{c|}{3.1}&\multicolumn{1}{c|}{2.1}&\multicolumn{1}{c|}{2.4}\\
\hline
\multicolumn{1}{|c|}{\multirow{1}{*}{Synovium}}&\multicolumn{1}{c|}{General}&\multicolumn{1}{c|}{81.8} & \multicolumn{1}{c|}{87.7} &  \multicolumn{1}{c|}{93.8} & \multicolumn{1}{c|}{89.5}\\
\hline
\end{tabular}
\caption{\label{tab:occ_map_tp_summary}Percentages of selected tissues identified as hotspots among 124 true positives detected by integrated MRI pipeline, stratified by OA severity.}
\end{table}

\begin{table}[ht]
\centering
\begin{tabular}{ |P{0.7cm}|P{0.3cm}|P{0.8cm}|P{1.9cm}|P{1.9cm}|P{1.9cm}|} 
\hline
\multicolumn{1}{|c|}{\multirow{2}{*}{\textbf{Tissue type}}}&\multicolumn{1}{c|}{\multirow{2}{*}{\textbf{Tissue}}}&\multicolumn{1}{c|}{\multirow{2}{*}{\begin{tabular}{@{}c@{}}\textbf{No OA} \\ \textbf{(n = 11)}\end{tabular}}}&\multicolumn{1}{c|}{\multirow{2}{*}{\begin{tabular}{@{}c@{}}\textbf{Moderate OA} \\ \textbf{(n = 65)}\end{tabular}}}&\multicolumn{1}{c|}{\multirow{2}{*}{\begin{tabular}{@{}c@{}}\textbf{Severe OA} \\ \textbf{(n = 48)}\end{tabular}}}&\multicolumn{1}{c|}{\multirow{2}{*}{\begin{tabular}{@{}c@{}}\textbf{Total} \\ \textbf{(n = 124)}\end{tabular}}}\\
&&&&&\\
\hline
\multicolumn{1}{|c|}{\multirow{3}{*}{Cartilage}}&\multicolumn{1}{c|}{TFJ medial}&\multicolumn{1}{c|}{100} & \multicolumn{1}{c|}{100} &  \multicolumn{1}{c|}{100} & \multicolumn{1}{c|}{100}\\
&\multicolumn{1}{c|}{TFJ lateral}&\multicolumn{1}{c|}{100} & \multicolumn{1}{c|}{100}&\multicolumn{1}{c|}{100}&\multicolumn{1}{c|}{100}\\
&\multicolumn{1}{c|}{PFJ}&\multicolumn{1}{c|}{18.2} & \multicolumn{1}{c|}{27.7}&\multicolumn{1}{c|}{62.5}&\multicolumn{1}{c|}{40.3}\\
\hline
\multicolumn{1}{|c|}{\multirow{4}{*}{Meniscus}}&\multicolumn{1}{c|}{Medial anterior}&\multicolumn{1}{c|}{100} & \multicolumn{1}{c|}{92.3} &  \multicolumn{1}{c|}{93.8} & \multicolumn{1}{c|}{93.5}\\
&\multicolumn{1}{c|}{Medial posterior}&\multicolumn{1}{c|}{100} & \multicolumn{1}{c|}{96.9}&\multicolumn{1}{c|}{83.3}&\multicolumn{1}{c|}{91.9}\\
&\multicolumn{1}{c|}{Lateral anterior}&\multicolumn{1}{c|}{90.9} & \multicolumn{1}{c|}{96.9}&\multicolumn{1}{c|}{97.9}&\multicolumn{1}{c|}{96.8}\\
&\multicolumn{1}{c|}{Lateral posterior}&\multicolumn{1}{c|}{100} & \multicolumn{1}{c|}{100}&\multicolumn{1}{c|}{91.7}&\multicolumn{1}{c|}{96.8}\\
\hline
\multicolumn{1}{|c|}{\multirow{3}{*}{Bone}}&\multicolumn{1}{c|}{TFJ medial}&\multicolumn{1}{c|}{100} & \multicolumn{1}{c|}{100} &  \multicolumn{1}{c|}{97.9} & \multicolumn{1}{c|}{99.2}\\
&\multicolumn{1}{c|}{TFJ lateral}&\multicolumn{1}{c|}{100} & \multicolumn{1}{c|}{100}&\multicolumn{1}{c|}{100}&\multicolumn{1}{c|}{100}\\
&\multicolumn{1}{c|}{PFJ}&\multicolumn{1}{c|}{9.1} & \multicolumn{1}{c|}{26.2}&\multicolumn{1}{c|}{56.3}&\multicolumn{1}{c|}{36.3}\\
\hline
\multicolumn{1}{|c|}{\multirow{3}{*}{Ligament}}&\multicolumn{1}{c|}{ACL}&\multicolumn{1}{c|}{100} & \multicolumn{1}{c|}{90.8} &  \multicolumn{1}{c|}{79.2} & \multicolumn{1}{c|}{87.1}\\
&\multicolumn{1}{c|}{PCL}&\multicolumn{1}{c|}{45.5} & \multicolumn{1}{c|}{63.1}&\multicolumn{1}{c|}{72.9}&\multicolumn{1}{c|}{65.3}\\
&\multicolumn{1}{c|}{Popliteal}&\multicolumn{1}{c|}{54.5} & \multicolumn{1}{c|}{46.2}&\multicolumn{1}{c|}{41.7}&\multicolumn{1}{c|}{45.2}\\
\hline
\multicolumn{1}{|c|}{\multirow{8}{*}{Tendon}}&\multicolumn{1}{c|}{Medial patellar retinaculum}&\multicolumn{1}{c|}{54.5} & \multicolumn{1}{c|}{66.2} &  \multicolumn{1}{c|}{87.5} & \multicolumn{1}{c|}{73.4}\\
&\multicolumn{1}{c|}{Lateral patellar retinaculum}&\multicolumn{1}{c|}{18.2} & \multicolumn{1}{c|}{23.1}&\multicolumn{1}{c|}{35.4}&\multicolumn{1}{c|}{27.4}\\
&\multicolumn{1}{c|}{Popliteal}&\multicolumn{1}{c|}{45.5} & \multicolumn{1}{c|}{33.8}&\multicolumn{1}{c|}{37.5}&\multicolumn{1}{c|}{36.3}\\
&\multicolumn{1}{c|}{Patellar}&\multicolumn{1}{c|}{9.1} & \multicolumn{1}{c|}{16.9}&\multicolumn{1}{c|}{18.8}&\multicolumn{1}{c|}{16.9}\\
&\multicolumn{1}{c|}{Gastrocnemius}&\multicolumn{1}{c|}{0.0} & \multicolumn{1}{c|}{9.2}&\multicolumn{1}{c|}{0.0}&\multicolumn{1}{c|}{4.8}\\
&\multicolumn{1}{c|}{Semimembranosus}&\multicolumn{1}{c|}{36.4} & \multicolumn{1}{c|}{26.2}&\multicolumn{1}{c|}{12.5}&\multicolumn{1}{c|}{21.8}\\
&\multicolumn{1}{c|}{Quadriceps}&\multicolumn{1}{c|}{0.0} & \multicolumn{1}{c|}{0.0}&\multicolumn{1}{c|}{6.3}&\multicolumn{1}{c|}{2.4}\\
&\multicolumn{1}{c|}{Gracilis}&\multicolumn{1}{c|}{0.0} & \multicolumn{1}{c|}{1.5}&\multicolumn{1}{c|}{0.0}&\multicolumn{1}{c|}{0.8}\\
\hline
\multicolumn{1}{|c|}{\multirow{1}{*}{Fat Pad}}&\multicolumn{1}{c|}{Hoffa}&\multicolumn{1}{c|}{81.8} & \multicolumn{1}{c|}{81.5} &  \multicolumn{1}{c|}{95.8} & \multicolumn{1}{c|}{87.1}\\
\hline
\multicolumn{1}{|c|}{\multirow{7}{*}{Muscle}}&\multicolumn{1}{c|}{Popliteus}&\multicolumn{1}{c|}{18.2} & \multicolumn{1}{c|}{15.4} &  \multicolumn{1}{c|}{10.4} & \multicolumn{1}{c|}{13.7}\\
&\multicolumn{1}{c|}{Vastus medialis}&\multicolumn{1}{c|}{0.0} & \multicolumn{1}{c|}{7.7}&\multicolumn{1}{c|}{16.7}&\multicolumn{1}{c|}{10.5}\\
&\multicolumn{1}{c|}{Gastrocnemius}&\multicolumn{1}{c|}{45.5} & \multicolumn{1}{c|}{29.2}&\multicolumn{1}{c|}{6.3}&\multicolumn{1}{c|}{21.8}\\
&\multicolumn{1}{c|}{Plantaris}&\multicolumn{1}{c|}{18.2} & \multicolumn{1}{c|}{15.4}&\multicolumn{1}{c|}{10.4}&\multicolumn{1}{c|}{13.7}\\
&\multicolumn{1}{c|}{Biceps femoris}&\multicolumn{1}{c|}{0.0} & \multicolumn{1}{c|}{1.5}&\multicolumn{1}{c|}{0.0}&\multicolumn{1}{c|}{0.8}\\
&\multicolumn{1}{c|}{Tibialis anterior}&\multicolumn{1}{c|}{9.1} & \multicolumn{1}{c|}{0.0}&\multicolumn{1}{c|}{0.0}&\multicolumn{1}{c|}{0.8}\\
&\multicolumn{1}{c|}{Semimembranosus}&\multicolumn{1}{c|}{27.3} & \multicolumn{1}{c|}{7.7}&\multicolumn{1}{c|}{2.1}&\multicolumn{1}{c|}{7.3}\\
\hline
\multicolumn{1}{|c|}{\multirow{1}{*}{Synovium}}&\multicolumn{1}{c|}{General}&\multicolumn{1}{c|}{90.9} & \multicolumn{1}{c|}{87.7} &  \multicolumn{1}{c|}{87.5} & \multicolumn{1}{c|}{87.9}\\
\hline
\end{tabular}
\caption{\label{tab:occ_map_tn_summary}Percentages of selected tissues identified as hotspots among 124 true negative controls detected by integrated MRI pipeline, stratified by OA severity.}
\end{table}

\end{document}